\documentstyle[aps,preprint,psfig]{revtex}
\tighten
\begin{document}
\draft
\author{J. P. Dewitz$^{a,b}$, Jian Chen$^c$, and W. H\"ubner$^b$}
\address{(a) Institut f\"ur Theoretische Physik, Freie Universit\"at
  Berlin, Arnimallee 14, D-14195 Berlin, Germany}
\address{(b) Max Planck Institut f\"ur Mikrostrukturphysik, Weinberg 2, D-06120 Halle/Saale, Germany}
\address{(c) Department of Physics, University of Florida, Gainesville, FL
  32611, USA.}
\date{\today}
\title{Nonlinear Magneto-Optics of Fe Monolayers from first
  principles: Structural dependence and spin-orbit coupling strength}
\maketitle

\begin{abstract}
We calculate the nonlinear magneto-optical response of free-standing fcc
(001), (110) and (111) oriented Fe monolayers. The bandstructures are
determined from first principles using a full-potential LAPW method
with the additional 
implementation of spin-orbit coupling. The variation of the spin-orbit
coupling strength and the nonlinear magneto-optical spectra upon layer
orientation are investigated. We find characteristic differences which
indicate an enhanced sensitivity of nonlinear magneto-optics to
surface orientation and variation of the in-plane lattice constants. In
particular the crossover from onedimensional stripe structures to
twodimensional films of (111) layers exhibits a clean signature in the
nonlinear Kerr-spectra and demonstrates the versatility of nonlinear
magneto-optics as a tool for in situ thin-film analysis.
\end{abstract}
%
\section{Introduction}
The Nonlinear Magneto-Optical Kerr effect (NOLIMOKE) is a unique
optical  tool
to analyze thin-film systems, since it is exclusively generated at
surfaces and interfaces, where the local inversion symmetry is
broken~\cite{shen}. Therefore is has attracted considerable interest
in recent
research on interface magnetism~\cite{shg1,shg2,shg3,shg4,stra,luce,dae,kiri,shg5}. In this
paper we investigate the dependence of the nonlinear magneto-optical
Kerr-spectra and their microscopic origin~\cite{agy} -- the spin-orbit coupling
(SOC) and the magnetic moment -- of Fe monolayers on structural changes. In particular the effects of different in-plane lattice
constants, different coordination and onedimensional stripe structures
are studied. The results were obtained using the full potential
linearized augmented plane wave (FLAPW) method WIEN95~\cite{wien95}
with the additionally performed implementation of spin-orbit coupling.

Two features are responsible for the strong interest in magnetic thin
films: (i) the magnetic properties strongly depend on structural changes
and (ii) the spin-orbit induced effects like magnetic anisotropy and giant
magnetoresistence (GMR) are much larger in low dimensional
systems. Whereas the magnetic anisotropy is in general larger in
twodimensional systems due to the reduced symmetry and thus a different
crystal potential, the change
of the magnetic moments is generated by the magneto-volume effect,
i.e. the dependence of the magnetic moment on the atomic
volume~\cite{Magnetovol1,Magnetovol2,Magnetovol3,Magnetovol4}. The latter
is affected by the differences between the equilibrium lattice
constants of substrate and overlayer and the induced overlayer
structures. One of the most striking examples is the Fe/Cu(001) system
which shows a rich magnetic phase diagram in the range from 1 to 11
monolayers~\cite{FeCu1,FeCu2,FeCu3}. Therein heavily distorted fcc
structures appear. The interlayer distances are relaxed (resulting in
a fct structure) and shifts of
the atomic positions in and perpendicular to the layer plane
occur~\cite{FeCu4}. Also the in-plane lattice
constants of consecutive layers are relaxed. For the Fe/Cu(001) system
experiments~\cite{stra} confirmed the sensitivity of NOLIMOKE on structural
changes of the top layer.

Since all these effects take place in configurations with only a few
monolayers it is possible to obtain them directly by {\em ab initio}
methods. A lot of calculations were done for freestanding and
supported monolayers, most of them aiming at the computation of magnetic
anisotropy. For 3$d$ transition metals the effects of
hybridization with the substrate~\cite{free}, different coordination
and $d$-band filling~\cite{wim,bluelet}, the stability of the
monolayer for different magnetic configurations, such as ferromagnetic,
antiferromagnetic or canted spin~\cite{blueapp,rossi,hafner}, and
superstructures~\cite{singh2} were calculated from first principles.

Since the theoretical prediction of the
sensitivity of nonlinear optics to surface
magnetism~\cite{pws,hue1,hue2} and the first
experiments~\cite{shg1,shg2} the applicability of NOLIMOKE to thin
film systems has been demonstrated for several
phenomena. Due to the fact that SHG is also generated at buried interfaces,
properties of different interfaces in multilayer systems could be
separated~\cite{rasi2,rasi3}. For that purpose it was important that the size of
the nonlinear magneto-optical effects, namely the nonlinear Kerr
rotation, is strongly increased compared to linear
optics~\cite{pusto2,pusto3}.  These
measurements also show a dependence on interface roughness~\cite{rasi1}.
 By direct comparison of linear
and nonlinear MOKE, changes of the magnetic
properties of the topmost layer during the growth process were
detected for the Co/Cu(001) system~\cite{rasi4}, since NOLIMOKE is
sensitive to the surface and interlayer only while
linear MOKE integrates over the
magnetism of all layers. Also quantum well states which occur in sandwich
structures could be resolved. This has been shown both
experimentally~\cite{kiri,straub1,straub2,rasi5} and
theoretically~\cite{luce}, by exploiting the fact that NOLIMOKE spectra
reflect characteristic features of the
bandstructures~\cite{pusto1}. Recent work
predicted that even the influence of surface antiferromagnetism on the optical signal can
be resolved by the nonlinear magneto-optical Kerr-effect
(NOLIMOKE)~\cite{dae}. This has already been shown before for SHG of the
antiferromagnetic non-inversion symmetric bulk Cr$_2$O$_3$~\cite{froe,gros}. 

One further potential of SHG, which to our knowledge has not been
applied to magnetic systems so far, is the strongly
enhanced sensitivity to submonolayer
coverages~\cite{shen2,rich}. Second harmonic generation 
by small particles is enhanced by local-field effects. In the case of
clusters deposited on a substrate this gives rise to signals for
particle sizes around 1~nm~\cite{trae}, which is far beyond
the resolution limit of linear optics. For spherical particles the
effects of local-field enhancement are well known by the linear Mie theory~\cite{mie}. Extensions
to nonlinear optics~\cite{stampf} show an enhanced sensitivity of the
size-dependent resonances compared to the linear
case~\cite{jpd}. In the case of 3$d$ transition-metal overlayers it
should be possible to resolve nanostructures of nm size with low
density by making use of the submonolayer coverage sensitivity of
SHG and the different in-plane symmetries of the nanostructures and
the substrate. From the experimental point of view the preparation of
nanostructures can now be achieved by state of the art techniques such
as molecular beam epitaxy varying the growth parameters (e.g. the
deposition rate or the temperature~\cite{kern1,kern2}). 

So far calculations of SHG generated by
metal surfaces are mainly restricted to simple and noble metals which
are well described by the model of a
free electron gas. These systems were intensely studied by Liebsch
and coworkers~\cite{lfund,ldft,lmetover}. They also calculated anisotropic
contributions~\cite{lanis1,lanis2,lanis3} and the influence of steps~\cite{lsteps1,lsteps2,lsteps3}
and obtain good agreement with experiment. Other authors studied the
change of the SHG yield in the presence of adsorbates on simple metal
surfaces within density functional theory~\cite{rebe1,rebe2,rebe3}. For these
nonmagnetic systems the intraband
transitions show stronger contributions than the interband
transition. Thus a better model for the screening effects is necessary, 
whereas in the case of transition metals the response is mainly due to
interband transitions. Then the intraband effects can be added by applying
a Drude model using experimental parameters~\cite{oppe1}. Calculations
of the linear magneto-optical Kerr-effect (MOKE) indicate that {\em ab initio} methods including spin-orbit
coupling and an highly accurate determination of the dipole transition
matrix elements are necessary to obtain magneto-optical
spectra which can be compared to experimental
values~\cite{oppe1,guo,ebertrev}. To some extent this was realized for
nonlinear magneto-optics by Pustogowa {\em et al.}~\cite{pusto4}. In their work the
Kerr spectra of Fe films with one to seven layers and the
dependence of the Kerr-spectra of a Fe(001) monolayer on the in-plane
lattice constant have been calculated by determining the electronic
bands within a full potential linear muffin tin orbital (FP-LMTO)
code. Spin-orbit coupling was treated within first-order perturbation
theory and the optical matrix elements were approximated as constants.

Here we will use the FLAPW method and go beyond this work within two
respects: different orientations of
Fe-monolayers ((001),(110),(111) of fcc) are investigated and, apart from the Kerr-spectra and the
magnetic moments, we focus on the spin-orbit coupling strength and
its structural dependence. Though microscopically both spin-orbit
coupling and spin-polarization are necessary to generate
magneto-optical response, spin-orbit coupling plays a special role, since the spin-orbit coupling
strength is directly proportional to the size of the magneto-optical
Kerr effect. This is known from studies of linear
MOKE~\cite{oppe2,mise}. Thus e.g.
the Kerr-rotation of an Fe/Pt system is much larger than that of an Fe
layer, since the large spin-orbit coupling of Pt contributes via
hybridization with the magnetic Fe-layer~\cite{guo,eber1}. This knowledge is important for applications in storage
technology, where magneto-optics is applied in a
configuration, where a perpendicular easy axis in combination with an
increased Kerr-rotation is preferred. In contrast to the spin-orbit
coupling the
dependence on the magnetic moment is rather complicated. Nevertheless
little is known about the spin-orbit coupling constants of thin film
systems contrary to their magnetic moments.

In our work the optical spectra are determined by using the same approximations
as in~\cite{pusto4}, i.e. the matrix elements are taken as constants and
the effects of spin-orbit coupling in the wavefunctions are treated within first-order perturbation theory. Since the spin-orbit
induced changes of the wavefunctions yield first-order effects~\cite{kittel} while
spin-orbit induced shifts of the eigenenergies give rise to second-order effects, we neglect spin-orbit coupling in the calculations of
the electronic bands, which are obtained from first principles. The
validity of this approach will be shown below.

By the choice of the investigated monolayers, we want to study several
aspects of structural changes. Firstly, we investigate the influence of relaxation of
the in-plane lattice constant, which is varied over a wide range for
the Fe(001) monolayer. Secondly, substrates of different orientations are
simulated by comparing the results for the Fe(001), Fe(110) and
Fe(111) monolayers, which also reveals the role of
coordination. These structures are deduced from the bulk fcc lattice. Two
lattice constants are considered, the lattice constant induced by Cu
fcc bulk and an even smaller value. Thirdly, the role of
nanostructuring is studied for regular arrays of stripes,
which can be created by viewing the closed monolayer as a regular
array of chains and then relaxing the distance between the
chains. Although this structure is rather artificial, it reveals the
effect of reducing the dimension of the layer in a second direction. Also
we compare our results with previous calculations.

Future work will address the calculation of the optical dipole matrix
elements to get the full information on the size of the NOLIMOKE spectra and to
exploit the symmetry properties of the 
systems, which will be of special interest in the calculation of
special nanostructures like triangular islands. This includes the
determination of the lateral resolution limit of nonlinear optics.

The paper is organized as follows: In Sec.~II, we will outline
the theory for the nonlinear magneto-optical response and our method
to calculate the spin-orbit coupling. Then the result part follows, which
is divided into three subsections (Secs.~III A, B, C), each for the comparison of different
characteristic changes of the structures. The paper ends with summary
and outlook (Sec.~IV).
%
\newpage
\section{Theory}
Within the electric-dipole approximation the polarization {\bf P} of the
medium can be expanded
in terms of the incident field {\bf E} as
\[P_i = \chi^{(1)}_{ij}E_j + \chi^{(2)}_{ijk}E_jE_k + \ldots\quad,\]
where $\chi^{(1)}$ and $\chi^{(2)}$ are the linear and second harmonic
susceptibilities~\cite{shen}.
We calculate the nonlinear magneto-optical response within the theoretical
framework introduced by H\"ubner and
Bennemann~\cite{hue1} and obtain the nonlinear susceptibility in the
electric dipole approximation as 
\begin{eqnarray}
  \label{eq:suszep1}
  \chi^{(2)}_{ijk}\left(2{\bf q},2\omega\right) = \frac{-{\rm
      i}e^3}{2q^3\Omega}\sum_{{\bf k},l,l',l''}\left\{\left<{\bf k}+2{\bf q},l''|i|{\bf
      k},l\right>\left<{\bf k},l|j|{\bf
      k+q},l'\right>\left<{\bf k+q},l'|k|{\bf k}+2{\bf
      q},l''\right>\right.\nonumber\\
\left.\times\frac{\frac{f(E_{{\bf k}+2{\bf
          q},l''})-f(E_{{\bf k}+{\bf
          q},l'})}{E_{{\bf k}+2{\bf
          q},l''}-E_{{\bf k}+{\bf
          q},l'}-\hbar\omega+i\hbar\alpha_1}-
    \frac{f(E_{{\bf k}+{\bf q},l'})- f(E_{{\bf
          k},l})}{E_{{\bf k}+{\bf q},l'}-
      E_{{\bf k},l}-\hbar\omega+i\hbar\alpha_1}}{E_{{\bf k}+
      2{\bf q},l''}-E_{{\bf k}l}-
    2\hbar\omega+i2\hbar\alpha_1}\right\}\;.
\end{eqnarray}
The indices $i$, $j$, $k$ run over $x$, $y$ and $z$.
In previous calculations~\cite{pusto4} the wave-functions and the band
energies were calculated neglecting spin-orbit coupling. Instead
spin-orbit coupling was taken as a perturbation and the product of the
three matrix elements were calculated using first order
perturbation theory to yield
\begin{equation}
\label{eq:3matpert}
\frac{\lambda_{\rm so}}{\hbar\omega}\left<{\bf k}+2{\bf q},l''|i|{\bf
      k},l\right>\left<{\bf k},l|j|{\bf
      k+q},l'\right>\left<{\bf k+q},l'|k|{\bf k}+2{\bf
      q},l''\right>
\end{equation}
where the wavefunctions and energies do not contain spin-orbit coupling
and the spin-orbit coupling constant is taken from the atomic value of
the spin-polarized $d$-bands. The matrix elements are approximated as
constants. This approach includes explicit inversion
symmetry breaking but makes it impossible to distinguish the
different elements of the tensor $\chi_{ijk}$. Nevertheless the resulting nonlinear susceptibility
\begin{equation}
\label{eq:suszep}
\chi^{(2)}(2{\bf q}_\parallel,2\omega,{\bf M})=\frac{C^3e^3{\bf
    q}_\parallel
  a}{\Omega}\frac{\lambda_{so}}{\hbar\omega}\sum_\sigma\sum_{{\bf
    k},l,l',l''} 
\frac{\frac{f(E_{{\bf k}+2{\bf
          q}_\parallel ,l''\sigma})-f(E_{{\bf k}+{\bf
          q}_\parallel ,l'\sigma})}{E_{{\bf k}+2{\bf
          q}_\parallel ,l''\sigma}-E_{{\bf k}+{\bf
          q}_\parallel ,l'\sigma}-\hbar\omega+i\hbar\alpha_1}-
    \frac{f(E_{{\bf k}+{\bf q}_\parallel ,l'\sigma})- f(E_{{\bf
          k},l\sigma})}{E_{{\bf k}+{\bf q}_\parallel ,l'\sigma}-
      E_{{\bf k},l\sigma}-\hbar\omega+i\hbar\alpha_1}}{E_{{\bf k}+
      2{\bf q}_\parallel ,l''\sigma}-E_{{\bf k}l\sigma}-
    2\hbar\omega+i2\hbar\alpha_1}\;,
\end{equation}
reflects the spectral dependence of
a magnetic tensor element, since spin-orbit coupling enters in first
order. Nonmagnetic tensor elements (and all even order tensor
elements) also consist of the zeroth order (and the corresponding
higher even orders) in spin-orbit coupling.
Thus they do not contribute to magneto-optics within first order and
yield larger values. Due to our approximations we add in Eq.~(\ref{eq:suszep}) a spin
index $\sigma$, drop the indices which specify the tensor elements and
add the factor $C^3$ originated by the approximate size of the matrix
elements. 

 The susceptibility is exclusively
built on interband transitions. We will use this approximation throughout this paper, since interband resonances
dominate the optical response of metallic systems. Thus in our case we
will call the dependence of $\omega^2{\rm Im}\chi^{(2)}(\omega)$ on
the photon energy NOLIMOKE ``spectra''. For details we refer to~\cite{pusto4}.

Calculations
including spin-orbit coupling will only affect the band-energies
$E_{\bf k}$, because the factor $\lambda_{\rm so}$ describes the effect
of SOC in the wavefunctions and the matrix elements, which are
included in $\chi^{(2)}$ as constants $C$, are not calculated
explicitly.

In this work the bandstructures are obtained from first
principles using the full potential linear augmented plane wave (FLAPW)
method WIEN95~\cite{wien95}. Additionally we implemented spin-orbit
coupling in a second variational step as described
e.g. by~\cite{free,daal2,singh}. 
After the self-consistent determination of the wavefunctions
and eigenenergies (quantities which are obtained self-consistently
without SOC are
marked by a suffix ``sc'' in the following) the Hamiltonian matrix is determined including
spin-orbit coupling
\[\sum_{ij}\left<\phi_{{\bf k}_i}^{\rm sc}\left|H^{\rm sc}+H_{\rm
      so}\right|\phi_{{\bf k}_j}^{\rm sc}\right> =
\sum_{ij}\epsilon(q)\rho_i\left(q\right)\left<\phi_{{\bf k}_i}^{\rm sc}|\phi_{{\bf k}_j}^{\rm
    sc}\right>\]
to obtain the eigenfunctions
\[\psi(q) = \sum_n\rho_n(q)\phi_{{\bf k}_n}\qquad q=1,2,\ldots\]
and the corresponding eigenenergies $\epsilon(q)$ shifted by
spin-orbit coupling ($q$ is the index of the eigenenergies and
$\rho_n(q)$ is the coefficient of the $n$-th basis-function in the
$q$-th eigenfunction). Here, spin-orbit coupling is not calculated self
consistently, especially the basis functions are not affected by
SOC. The procedure is known to yield good agreement with exact results~\cite{daal}.

To determine the spin-orbit part of the Hamiltonian the basis
functions of the FLAPW method have to be taken into account. The basis
set consists of the standard basis functions
\begin{equation}
  \label{eq:basis}
\phi_{{\bf k}_i} = \left\{
\begin{array}{lc}\sum\limits_{lm}\left[A_{lm}\left({\bf
        k}_i\right)u_l\left(r,E_l\right)+B_{lm}\left({\bf
        k}_i\right)\dot{u}_l\left(r,E_l\right)\right]Y_{lm}
  & r<{\rm R}_{\rm mt}
  \\[1ex] \frac{1}{\sqrt{\omega}}e^{{\rm i}{\bf k}_i{\bf r}} & r>{\rm R}_{\rm mt
}\end{array}
\right.
\end{equation}
and the so called {\em local orbitals}, which are introduced to
describe the low lying semi-core states~\cite{singh,los}
\begin{equation}
  \label{eq:LOs}
  \phi_{{\bf k}_i}^{\rm LO} = \left\{
\begin{array}{lc}\sum\limits_{lm}\left[A_{lm}\left({\bf
        k}_i\right)u_l\left(r,E_l\right)+B_{lm}\left({\bf
        k}_i\right)\dot{u}_l\left(r,E_l\right)+C_{lm}\left({\bf k}_i\right)\hat{u}_l\left(r,E_2\right)\right]Y_{lm}
  & r<{\rm R}_{\rm mt}
  \\[1ex] 0 & r>{\rm
    R}_{\rm mt}\end{array}
\right.\;.
\end{equation}
and are included for all $l$-values for which semi-core states appear
($R_{\rm mt}$ is the muffin tin radius). The radial functions are
obtained from the Schr\"odinger equation 
\begin{equation}
  \label{eq:udef}
  \left[ -\frac{2}{r}\frac{\partial}{\partial r} -
  \frac{\partial^2}{\partial r^2} + \frac{l(l+1)}{r^2} + V^\sigma(r) \right] u_l^\sigma(r) = E_l^\sigma u_l^\sigma(r)\;,
\end{equation}
where the localization energies $E_l^\sigma$ are chosen to be at the
center of the band. 

The spin-orbit operator 
\[H_{\rm so}=\frac{\alpha^2}{2}{\bf
  s}\cdot\left(\vec{\nabla}V\times{\bf p}\right)\]
($\alpha$ is the fine structure constant) is applied in the spherical approximation: 
  \[
   \vec{\nabla}V = \frac{\bf r}{r}\frac{\partial V}{\partial
    r}\]
\[\frac{\partial V}{\partial r}\equiv 0,\qquad r>R_{\rm mt}\;,
  \]
since the gradient of the potential yield its largest contributions
near the core, where the potential is almost spherical.
This yields
\[H_{\rm so}=\frac{\alpha^2}{2}{\bf s}\cdot\left({\bf r}\times{\bf p}\right)\frac{1}{r}\frac{\partial V}{\partial r} =
\frac{\alpha^2}{2}{\bf s}\cdot{\bf L}\frac{1}{r}\frac{\partial
  V}{\partial r}.\]
The spin-orbit matrix elements
\begin{equation}
\label{eq:socmat}
\left<\phi_{{\bf k}_i}^{\rm sc}\left|H_{\rm so}\right|\phi_{{\bf
      k}_j}^{\rm sc}\right>=\int_{r<{\rm R_{mt}}} d{\bf r}\;\phi^{{\rm sc}*}_{{\bf
    k}_i}\left(\frac{\alpha^2}{2}{\bf s}\cdot{\bf
    L}\frac{1}{r}\frac{\partial V}{\partial r}\right)\phi^{\rm
    sc}_{{\bf k}_j}
\end{equation}
\[\left(\int_{r>{\rm R_{mt}}}d{\bf r}\;\phi^{{\rm sc}*}_{{\bf
    k}_i}H_{\rm so}\phi^{\rm sc}_{{\bf k}_j}\equiv 0\right)\]
are calculated by separating the angular and radial parts. This yields
\begin{eqnarray*}
 =\sum_{lmm'}\left\{\right.&\lambda_{uu}^l& A_{lm}^*\left({\bf
        k}_i\right)A_{lm'}\left({\bf k}_j\right)+ \\
&\lambda_{\dot{u}\dot{u}}^l& B_{lm}^*\left({\bf
        k}_i\right)B_{lm'}\left({\bf k}_j\right)+\\
&\lambda_{u\dot{u}}^l&\left.\left[A_{lm}^*\left({\bf
        k}_i\right)B_{lm'}\left({\bf k}_j\right)+B_{lm}^*\left({\bf
        k}_i\right)A_{lm'}\left({\bf k}_j\right)\right]\right\}\\
&\cdot&\left<\sigma\left|\int d\Omega\; Y_{lm}^*\left(\hat{r}\right){\bf
      s}\cdot{\bf L}Y_{lm'}\left(\hat{r}\right)\right|\sigma'\right>
\end{eqnarray*}
with the spin-orbit coupling constants
\begin{eqnarray}
\label{eq:socc}
  \lambda_{uu}^l&\equiv& \frac{\alpha^2}{2}\int_{r<{\rm R}_{\rm mt}}dr\;u_l^\sigma\left(r\right)r\frac{\partial V_{\sigma'}}{\partial
    r}u_l^{\sigma'}\left(r\right)\nonumber\\ 
\lambda_{u\dot{u}}^l&\equiv& \frac{\alpha^2}{2}\int_{r<{\rm R}_{\rm mt}}dr\;u_l^\sigma\left(r\right)r\frac{\partial V_{\sigma'}}{\partial
    r}\dot{u}_l^{\sigma'}\left(r\right)\\
\lambda_{\dot{u}\dot{u}}^l&\equiv& \frac{\alpha^2}{2}\int_{r<{\rm
    R}_{\rm mt}}dr\;\dot{u}_l^\sigma\left(r\right)r\frac{\partial V_{\sigma'}}{\partial
  r}\dot{u}_l^{\sigma'}\left(r\right)\nonumber
\end{eqnarray}
and the spin-orbit coupling constants including local orbital
functions:
\begin{eqnarray}
\label{eq:socclo}
  \lambda_{\hat{u}\hat{u}}^l&\equiv& \frac{\alpha^2}{2}\int_{r<{\rm R}_{\rm mt}}dr\;\hat{u}_l^\sigma\left(r\right)r\frac{\partial V_{\sigma'}}{\partial
    r}\hat{u}_l^{\sigma'}\left(r\right)\nonumber\\ 
\lambda_{\hat{u}u}^l&\equiv& \frac{\alpha^2}{2}\int_{r<{\rm R}_{\rm mt}}dr\;\hat{u}_l^\sigma\left(r\right)r\frac{\partial V_{\sigma'}}{\partial
    r}u_l^{\sigma'}\left(r\right)\\
\lambda_{\hat{u}\dot{u}}^l&\equiv& \frac{\alpha^2}{2}\int_{r<{\rm
    R}_{\rm mt}}dr\;\hat{u}_l^\sigma\left(r\right)r\frac{\partial V_{\sigma'}}{\partial
  r}\dot{u}_l^{\sigma'}\left(r\right)\nonumber
\end{eqnarray}
Thus we get three (six when local orbitals are involved) spin-orbit
coupling constants for one $l$-value which are formed by a radial
integral over the radial part of the basis functions and the radial
derivative of the potential. Furthermore one has to take into account
that spin-orbit coupling mixes the spins, thus the Hamilton-matrix gets
off-diagonal elements within the space of the spin $\uparrow$ and $\downarrow$ basis
functions. Spin is not a good quantum number anymore and the wavefunctions
consist of both spin $\uparrow$ and spin $\downarrow$ contributions. The spin-orbit
matrix elements Eq.~(\ref{eq:socmat}) get then additional spin indices $\sigma$
and $\sigma'$.
\begin{equation}
  \label{eq:socmats}
\left<\phi_{{\bf k}_i,\sigma}^{\rm sc}\left|H_{\rm so}\right|\phi_{{\bf
      k}_j,\sigma'}^{\rm sc}\right>=\int_{r<{\rm R_{mt}}} d{\bf r}\;\phi^{{\rm sc}*}_{{\bf
    k}_i,\sigma}\left(\frac{\alpha^2}{2}{\bf s}\cdot{\bf
    L}\frac{1}{r}\frac{\partial V_{\sigma'}}{\partial r}\right)\phi^{\rm
    sc}_{{\bf k}_j,\sigma'}
\end{equation}
Therein the spin-index of the potential is equal to the spin index
of the basis function on the right, since the spin-orbit operator acts
on it. The fact that the potentials are different for the spins, but
the basis functions are not, leads to the requirement to make the
matrix explicitly hermitian, since the spin-orbit operator
is. This affects only the spin-flip matrix elements. 
\newpage
\section{Results}
In Sec.~III. A we will simulate the effect of
lattice relaxation. This can be achieved experimentally by different
substrates, assuming pseudomorphic growth. We will show NOLIMOKE
spectra of free-standing Fe(001) monolayers with in-plane lattice
constants varied from $a=2.4$~\AA, which is slightly below the value
of the nearest-neighbor distance in Cu fcc bulk $a$=2.56~\AA, to $a=2.76$ \AA, which is close to the
nearest-neighbor distance of Fe bcc bulk. For the comparison of the trends of
the magnetic moments and
the spin-orbit coupling constants we extend the range of lattice
constants from 2.22~\AA~to 3.18~\AA, the latter corresponding
approximately to the value of bcc W. In Sec.~III.B the same
quantities are shown for Fe monolayers with different structures,
i.e. the fcc (111), (001) and (110), which are schematically displayed
in Fig.~\ref{fig:figgeo}. The structures are studied for the Cu fcc
nearest-neighbor distance $a$=2.56~\AA~and $a=2.4$~\AA$\;$. It should be
possible to get a measure of the structural changes from
the NOLIMOKE spectra. 
In Sec.~III.C we will show the influence of nanostructuring on the
NOLIMOKE spectra by analyzing stripe structures as indicated in
Fig.~\ref{fig:figchains}. The Fe(111) monolayer can be interpreted as an
array of ``zig-zag''-stripes. To reduce the dimension of the structure
we vary the distance of the stripes, which is indicated by $d$,
where in the case of $d=h$ the layer is equal to the (111)
structure. 
\subsection{Fe(001) monolayers}
Fig.~\ref{fig:fe001noli} shows the NOLIMOKE spectra $\omega^2{\rm
  Im}\chi^{(2)}(\omega)$ of the Fe(001) monolayer
as a function on the in-plane lattice constant. 
As analyzed by Pustogowa {\em et al.}~\cite{pustosurf} within a tight
binding scheme the first maximum and the zero are mainly due to
features of the $d$ bands, whereas for higher photon energies the role of the
$s$-$p$-bands is more dominant. In particular they showed that the position of the
zero is a measure for the $d$-band width and the height of the
maximum is proportional to the magnetic moment. Thus the $d$
bands generate the features of the spectra in the optical region. In our case the zeros show a clear dependence on the lattice
constant. The positions shift to lower energies with increasing
lattice constant. Since this point characterizes the $d$-bandwidth,
the bandwidth is reduced upon lattice expansion. From the bandstructure it can also be
seen that bands above the range of visible frequencies are shifted to lower energies with
increasing lattice constant, which generates the different slopes in the
high-energy part of the spectra. The height of the maxima starts to
increase with the spin-polarization for lattice constants from
2.4~\AA~to 2.58~\AA$\;$. For the larger lattice constants
($a$=2.67 and 2.76~\AA) there are no more significant changes of the
peak height, as can be seen in the inset of Fig.~\ref{fig:fe001noli}. However, the position of the
peaks is shifted in proportion to the value of the magnetic moments to
lower energy values. The dependence of the maximum on the magnetic
  moments agrees with previous works~\cite{oppe2,mise}, where no clear
  dependence of the linear magneto-optical response on the size of the
  magnetic moments was found. Additionally, from a tight binding calculation, Pustogowa {\em et
  al.}~\cite{pustosurf} found a linear dependence of the maximum for
  magnetic moments between 0 and 2.5~$\mu_B$, but a similar behavior
  for moments between 2.5~$\mu_B$ and 3.4~$\mu_B$. The
difference should reflect that in both calculations the magnetic
moments are changed by different mechanisms. Whereas in the tight
binding calculations the magnetic moments were affected by changes of
the exchange coupling constant $J$, in our case the magnetic moments
are varied by changing the lattice constant, which does not only shift
the relative positions of the $d$ subbands, but also their width. 

The values of the magnetic moments increase with increasing
lattice constant. This is shown in Fig.~\ref{fig:fe001muel}, where the
size of the magnetic moments (filled circles) is plotted as a function
of the in-plane lattice constant in units of
$\mu_{\rm B}$. If interpolated our results agree very well with
calculations by Wang {\em et al.}~\cite{wang}, who obtained
3.04$\mu_{\rm B}$ for a lattice constant of $a=2.56$~\AA, and with
results by Gay and Richter~\cite{gay}, who obtained $3.20\mu_{\rm B}$
for a lattice constant of $a=2.88$~\AA~compared to our results of
3.08~$\mu_B$ for $a=2.56$~\AA~and 
3.24$\mu_{\rm B}$ for $a=2.88$~\AA, respectively. Nevertheless one has to
keep in mind that the error in the magnetic moments is around $\pm
5$\% due to the chosen accuracy in our calculations.

In Fig.~\ref{fig:fe001muel} we also compare the values of the magnetic
moments directly with the spin-orbit coupling constants
$\lambda^{l=2}_{uu}$~$\uparrow\uparrow$ and
$\lambda^{l=2}_{uu}$~$\downarrow\downarrow$ defined in Eq.~(\ref{eq:socc}), i.e. the
spin-orbit coupling constants for bands with $d$ character for
$\uparrow$ and $\downarrow$ spin. These constants are the important
ones for magneto-optics since the $d$-bands exhibit the magnetic
moments. Combinations of the radial functions other than $(u_l,u_l)$
are of less interest since the radial dependence of wavefunctions is
mainly described by the $u_l$-functions. The plot shows two important
properties of the coupling constants: (i) The values increase with
decreasing lattice constants and (ii) the difference between the coupling
constants of $\uparrow$ and $\downarrow$ spin show a clear dependence
on the magnetic moments. 
Inspection of the potentials for the different lattice constants
shows that the size of the spin-orbit coupling constants is not
directly governed by changes of the potential, i.e. the derivative of
the potential shows no changes near the core, where the largest values
of the derivative occur. As a consequence, the changes of the constants
must be induced by changes of the radial functions. This is shown in
Fig.~\ref{fig:fe001lamb}, where the square of the function $u_l(r)$
and the integrand $u_ldV/dr u_l$ defined in Eq.~(\ref{eq:socc}) are
plotted as a function of the radial distance. The insets show that the
increase of the maximum of the integrand, which causes the changes of
the coupling constants with decreasing lattice constants, is
proportional to the changes of the square of $u_l$ (the maxima of both quantities are normalized
to one). Thus the changes of the coupling constants are caused by
changes of the potential near the muffin tin radius R$_{\rm mt}$, which alters the
probability of the maximum of the radial functions also close to the nuclei. In addition the
dependence of the differences between the $\uparrow\uparrow$ and
$\downarrow\downarrow$ coupling constants on the magnetic moments
reveals the variation of the potential with different spin-subband occupation
and by changing the radial functions via Eq.~(\ref{eq:udef}). For even
larger values of the lattice constants the coupling
constants should reach the atomic value, which is approximately
50~meV. In the case of Ge, the spin-orbit splitting of the 4$p$
electrons in the solid is 0.43~eV~\cite{hybertsen} at the $\Gamma$
point, a 30\% enhancement to the spin-orbit splitting of 0.21~eV in
the Ge atom. By comparing the wavefunctions and potentials in the
solid and the atom, we find that this increase in the spin-orbit
coupling strength in the solid in Ge is caused by a quite different
reason. Because of the covalent bond the charge distribution is not
only enhanced between the atoms but also near the core. Daalderop {\em et al}~\cite{daal2}
obtained the coupling constants of bulk Fe and find
a much larger difference between the coupling constants for
$\uparrow$ and $\downarrow$ spin and also the values differ slightly. Since
they used a LMTO code the differences should mainly be due to the
different definitions of the coupling constants resulting from the different
basis sets used. 

In Tab.~\ref{tab:fe001} the values of the additional spin-orbit
coupling constants for $l=1$ and 2 and within the combinations of the
radial functions $(u_l,u_l)$ and $(\dot{u}_l,\dot{u}_l)$ are
listed. The dependence of the coupling constants $\lambda_{uu}^l$ with
$l=1$ on the lattice constant differs significantly from the values for
$l=2$. Their changes are much more pronounced, namely the values
decrease by about 40\% rather than by only 5\% for $l=2$. Also a
spin-polarization appears only for the largest value of the lattice
constants whereas it slightly changes for $l=2$. These properties
reveal that the $p$ states are much more influenced by the changed
binding characteristics. Though in the case of Fe monolayers the $p$ band is not occupied and
its center is tens of eV above the Fermi-level, $p$-states could
become accessible by optical excitations via hybridization with low
lying $s$- and $p$-bands of an appropriate substrate such as Mo or W.
In these systems the large values of
the SOC-constants  could, in particular in the case of
pump-probe femtosecond spin dynamic experiments, significantly affect
spin-orbit induced spin-flip contributions by strong excitations.

In principle contributions related to the radial functions $\dot{u}_l$ should not contribute
significantly since the values of the coefficients $B_{lm}$ are in
general much smaller than $A_{lm}$, nevertheless the changes of the
coupling constants $\lambda_{\dot{u}\dot{u}}^l$ should reflect some
features of the shape of the bands. Since the values for $l=2$ show no
spin-polarization, the shape of the subbands should be nearly equal, also the derivative
increases quite strongly with the lattice constants indicating
narrower bands. For $l=1$ the increase is even stronger in agreement
with the values for $\lambda_{uu}^{l=1}$, the smaller values compared
to $l=2$ reflect stronger dispersed bands.

A direct comparison of the spectra of the Fe(001) monolayer for
$a=2.76$~\AA~obtained within the FLMTO method~\cite{upmono} and our
FLAPW method in Fig.~\ref{fig:fe001upjd} shows good agreement in the
region of low photon energies. The position of the maximum is near 1.5~eV in
both cases and the energy where the susceptibility crosses zero is
3~eV. In both calculations the same model for the nonlinear
magneto-optical susceptibility was applied. Thus the differences in the
region of higher photon energies should be an effect of the different
{\em ab initio} methods and in this special case due to the different
basis sets. In the LMTO-method the number of basis functions is much
smaller than in the FLAPW method which leads to a lack of bands high
above the Fermi-level unless the calculations are performed for
several localization energies. This is in agreement with the fact that
for spectra which are based on the Fe bulk bandstructure we find no
significant differences in both methods also for high photon energies.

Our results obtained for the changed lattice constants agree very well
with the results by Pustogowa {\em et al.}~\cite{pusto4} for the same
system. Since they calculated spectra for lattice constants larger
than 2.76~\AA~their changes of the zero are smaller due to
the nonlinear dependence of the shape of the bands on the lattice
constants, which is also reflected by the dependence of the magnetic
moments on the lattice constants in Fig.~\ref{fig:fe001muel}. In
contrast to their calculations in our case the position of the maximum
shows a clear dependence on the lattice constant.

The optical spectra also depend on the type of approximations applied to the calculations of the electronic bands. This can be seen in
Fig.~\ref{fig:gganoli}, where the spectrum of a Fe(001)
monolayer with $a=2.76$~\AA~is calculated using different approximations for the exchange
correlation potential. We compare the generalized
gradient approximation (GGA), which is used for all calculations
throughout this work, in the parameterization by Perdew {\em et
  al.}~\cite{ggapw} with the local spin density approximation (LSDA) in the parameterization of Perdew and Wang~\cite{lsdapw}. Since GGA corrects for
overbinding, the bandwidth should be lowered and thus the
zero should be at lower energy. In our case the opposite
behavior occurs. The LSDA values are lower in energy. The spectra obtained with
the different LSDA approximations show no strong deviations. GGA
yields a slightly higher magnetic moment which can be responsible for
the higher value of the maximum and the larger value of the zero. In general it is expected that
spin-orbit coupling counteracts GGA, since the
bandwidth increases by spin-orbit induced shifts. But in the case of
the NOLIMOKE spectra the effects of spin-orbit coupling which enters
the spectra via the bandstructure are negligible as can be seen in
Fig.~\ref{fig:fe001soc}, where the NOLIMOKE spectrum of Fe(110) is plotted
both with and without SOC effects on the bandstructure. This
reflects that spin-orbit induced changes of the eigenenergies only
contribute in second order to the spectrum~\cite{kittel}.

\subsection{Fe(001), (110) und (111) monolayers}
In Figs.~\ref{fig:strucCunoli} and~\ref{fig:strucnoli} the
NOLIMOKE spectra of the Fe(001), (110) and (111) monolayers are
compared for the nearest-neighbor distance of Cu fcc bulk,
$a$=2.56~\AA, and for $a$=2.4~\AA$\;$. Since the nearest-neighbor
distance is equal in the
different structures, the changes reveal the
different coordination, which is
six in the hexagonal (111) layer, four in the square lattice (001)
and two in the rectangular lattice (110). The different coordination
determines the area of the twodimensional (see Fig.~\ref{fig:figgeo})
unit cell containing one atom to $\sqrt{2}a^2$
for the (110) structure, $a^2$ for (001) and $\sqrt{3}/2a^2$ for
(111). The next nearest neighbor distance is $\sqrt{2}a$ in the (110),
$2a$ in the (001) and $\sqrt{3}a$ in the (111) layer.

For both nearest-neighbor distances
it can be seen that the lattice with the lowest coordination shows the
smoothest spectra, whereas for highest coordination the most
complicated structure appears. This is a general aspect of
coordination and can also be seen in the bandstructures~\cite{coord}. Between 0~eV
and the zero around 3 to 4~eV the (110) spectrum has a sinusoidal shape, in the (001) spectra
first a maximum followed by a shoulder appears, whereas in the (111)
case a dominant maximum is surrounded by two shoulders. The
differences are much more pronounced for the smaller nearest-neighbor distance,
in the case of $a$=2.56~\AA~the spectra are closer, as can be seen e.g.
at the zero point, and it is more difficult to define a maximum. Comparing the positions of the zero for both lattice
constants one notices that the shifts are larger for higher
coordination. Whereas the zero remains more or less constant in
the case of Fe(110), it is shifted to lower energies by approx. 0.2
eV for Fe(001) and 0.4~eV for Fe(111). Thus one can say that
the dependence of the spectra on the lattice constants is proportional
to the coordination. 

Roughly the same holds for the magnetic moments in
Figs.~\ref{fig:strucCumuel} and~\ref{fig:strucmuel}. There the
magnetic moments and the spin-orbit coupling constants
$\lambda_{uu}^{l=2}$~$\uparrow\uparrow$ and
$\lambda_{uu}^{l=2}$~$\downarrow\downarrow$ are plotted for three
different coordination numbers corresponding to (111), (001) and (110)
layers. The magnetic moments of the Fe(110) layers change by
only approx. 0.1~$\mu_B$ compared to approx. 0.25~$\mu_B$ for the
(111) and (001) monolayers. As expected the values of the magnetic
moments increase with lowered coordination. 
Comparison of the spin-orbit coupling constants show that the
changes induced by the different coordination are quite small compared
to the changes induced by different nearest-neighbor distances. Thus one can
say that in a first approximation the values of the spin-orbit coupling
constants depend on the nearest-neighbor distance and remain constant
for different coordination. 
The values of the coupling constants with $l=1$ and the combination of
the radial functions ($\dot{u}_l,\dot{u}_l$) confirm this since they show
no significant changes with the coordination as 
can be seen from Tabs.~\ref{tab:strucCu} and~\ref{tab:struc}.

\subsection{Stripe structures}
The NOLIMOKE spectra of the stripe structures of
Fig.~\ref{fig:figchains} are plotted in
Fig.~\ref{fig:stripesnoli} for different distances $d$ of the stripes
 in comparison to the spectrum of the Fe(111) monolayer, which can
be interpreted as a stripe structure with distance $d=h$ (see
Fig.~\ref{fig:figchains}). The spectra show no
behavior which can be easily
interpreted in terms of the bandwidth, corresponding to a zero
point in the spectra, or the magnetic moment, corresponding to a
maximum at a certain position in the low-energy-regime. Only the more
complicated structure of the spectra for $d>h$ is clear from the
lifting of degeneracies in the bandstructure, which results from the
breaking of symmetry. Thus, the number of bands increases for $d>h$, since there are two
nonequivalent atoms in the unit cell, compared to one for
Fe(111). Since the
bandstructures of the different stripe structures show no pronounced
overall changes, the differences in the spectra should be an effect of
the details of the bands.

If one first
neglects the spectra for the stripes with the largest distance
(long-dashed line) the behavior is quite regular in the sense that the
maximum value of the spectra decreases with increasing distance and
that the zero point shifts to lower energies and reaches zero for
$d=h+1.04$~\AA$\;$. By drastically increasing the ``interstripe'' distance to $d=h+1.78$~\AA~the
NOLIMOKE spectra differ from these trends and exhibit a shape which is
similar to the spectra of the closed layers in
Fig.~\ref{fig:fe001noli},~\ref{fig:strucCunoli} and~\ref{fig:strucnoli}. This can be interpreted as an oscillatory
behavior of the electronic structure with the distance. The change of
the trends for the distance of $d=h+1.78$~\AA~can also be observed in
Fig.~\ref{fig:stripesmuel} for the spin-orbit coupling constants and
more or less also for the magnetic moments. Since the difference
of their values for $d=h+1.04$~\AA~and $d=h+1.78$~\AA~is very small, $\mu$ remains essentially constant. 

The relatively small changes of the spin-orbit coupling constants with
$l=2$ shown in Fig.~\ref{fig:stripesmuel} and Tab.~\ref{tab:stripes}
imply the same interpretation as in Sec.~III.B for the layers with
different coordination. In a first approximation the values of the
coupling constants remain unchanged and thus their
values depend mainly on the nearest-neighbor distance, which is fixed here
 by the constant structure of the
isolated stripes. Thus, for tailoring the SOC-constants and the
magnetic moments, the choice of the substrate will be much more
efficient than nanostructuring while keeping the nearest-neighbor
distance constant. Nevertheless nanostructuring can still have a strong
effect on dynamical properties of the spin and magnetic moments, for
which the size of the SOC-constants is quite important. For certain
island sizes a phase transition from ferro- to superparagnetism will
occur and affect the spin dynamics.

The behavior of the coupling constants with $l=1$, which are also
listed in Tab.~\ref{tab:stripes}, show one difference. The values are
also nearly constant for distances larger than $h$, but they show a clear
increase, when the distance changes from $d=h$ to $d=h+0.46$~\AA,
i.e. at the onset of the reduction of the symmetry. Thus the
SOC-constants of the $p$-bands
are not only much larger but also much more sensitive to the bond characteristics than the $d$-bands.

Clearly further investigations are necessary to understand the
strong changes of the NOLIMOKE spectra for the different stripe distances.
\newpage
\section{Summary and Outlook}
We presented results for the structural dependence of the nonlinear
magneto-optics of Fe monolayers. Since in our theory the optical
dipole-transition matrix elements are approximated as constants, we
cannot analyze the symmetry properties of the nonlinear susceptibility tensor. Instead we focused on the spectral dependence
of a magnetic tensor element, the
magnetic moments, and the spin-orbit coupling constants, the latter
two reflecting the microscopic origin of magneto-optics.

In the case of the Fe(001) monolayer-spectrum the characteristic
features like the position of the first zero, which is related to the $d$-band
width, and the position of the maxima are shifted to lower energies
with decreasing lattice constant. The changes are stronger for smaller
lattice constants, which also holds for the magnetic moments and
the spin-orbit coupling constants. The values of the maxima, which
should be related to the magnetic moments show no clear trend. The
spectra of layers with different coordination numbers show characteristic
differences in the shape as well as in the position of the maximum and
the first zero. The differences are more pronounced for smaller lattice
constants. Reducing the dimensionality of the monolayers, simulated by
onedimensional stripes with different ``interstripe''-distances,
results in dramatic changes of the spectra. Their shapes show no
similarities with the monolayer spectra any more. In contrast the
values of the spin-orbit coupling constants depend in a first
approximation only on the nearest-neighbor distance. This was shown
for both the Fe monolayers with different coordination number and the
onedimensional stripe structures. 

For the Fe(001) monolayers the
SOC-constants show the opposite behavior as the magnetic moments, they
increase with decreasing lattice constants. As an important result the
difference between the coupling constants for $\uparrow$ and
$\downarrow$ spin is proportional to the magnetic moments, caused by
the dependence of the potentials on the occupation of the
subbands. The values of the magnetic moments show results well-known
for itinerant ferromagnets. Increasing the lattice constants or decreasing the
coordination enhances their values. The same holds for increasing the
distances of stripes in the quasi-onedimensional structures.

Our results clearly show the strong dependence of the NOLIMOKE spectra
on structural changes and also indicate that the spectral dependence
of the magneto-optical response is a valuable source of
information on the structure of the investigated system.

Future work will address the completion of our {\em ab initio} theory
by the optical transition dipole matrix elements, which is of major
importance not only for the determination of absolute signal values but also
for the study of structural dependencies, since the susceptibility
tensor reflects the symmetry of the system. Also we will investigate
structures with further reduced dimensionality, i.e. zerodimensional
islands, and apply nonlinear magneto-optics to antiferromagnets.

We acknowledge financial support by Deutsche Forschungsgemeinschaft
through Sfb 290 and by TMR Network NOMOKE contract No. FMRX-CT96-0015.
%

%
\begin{figure}[htbp]
  \caption{Geometry of the different layers investigated in this
    paper. The twodimensional unit cell containing one atom is shown,
    deduced from the fcc lattice in the (110), (001) and (111) direction.}
  \label{fig:figgeo}
  \caption{Chains built by stretching the Fe(111) monolayer as
    indicated. The unit cell used for the bandstructure calculation is
    indicated by the solid rectangle. To simulate the chains the distance $d$
    is increased compared to $h$. In the case of $d=h$
    the layer is equal to the Fe(111) monolayer.}
  \label{fig:figchains}
  \caption{NOLIMOKE spectra of Fe(001) monolayers with the lattice
    constant varying from $a=2.4$~\AA~to $a=2.76$~\AA$\;$.}
  \label{fig:fe001noli}
  \caption{Magnetic moments and spin-orbit coupling constants for
    Fe(001) monolayers as a function of the lattice
    constant. $\lambda_{uu}^{l=2}$ up
    and $\lambda_{uu}^{l=2}$ dn denote the SOC-constants obtained from the
    matrixelements within the spin-combinations $\uparrow\uparrow$ and
    $\downarrow\downarrow$.}
  \label{fig:fe001muel}
  \caption{Square of the radial function $u_2(r)$ and integrand
    $u_2(r)^2\cdot dV/dr$ as a function of the radius. In both cases
    the highest values are normalized to unity. The scale of the
    y-axes in the insets are equal, which indicates that changes in
    the integrand are directly caused by the radial functions.}
  \label{fig:fe001lamb}
  \caption{NOLIMOKE spectra in the case of a Fe(001) monolayer
    obtained by Pustogowa {\em et al.}~[75] using an FP-LMTO
    code and by the present authors using the FLAPW method.}
  \label{fig:fe001upjd}
  \caption{\protect Comparison of the NOLIMOKE spectra of a Fe(001)
    monolayer, obtained by using
    GGA~[76], LSDA with the parameterization of~[77].}
  \label{fig:gganoli}
  \caption{Comparison of the NOLIMOKE spectra obtained for the Fe(110)
    monolayer without spin-orbit coupling and with band shifts induced
    by SOC. The effect of SOC in the wavefunctions
    via the optical dipole matrix elements has not been taken into
    consideration.} 
  \label{fig:fe001soc}
  \caption{NOLIMOKE spectra as a function of the photon energy of the
    fundamental light for Fe(001), Fe(110) and Fe(111) monolayers
    using the Cu fcc lattice constant $a$=2.56~\AA$\;$.}
  \label{fig:strucCunoli}
  \caption{NOLIMOKE spectra as a function of the photon energy of the
    fundamental light for Fe(001), Fe(110) and Fe(111) monolayers
    using the lattice constant $a$=2.4~\AA$\;$.}
  \label{fig:strucnoli}
  \caption{Magnetic moments and spin-orbit coupling constants
    $\lambda_{uu}^{l=2}$ for the spin-combinations $\uparrow\uparrow$
    and $\downarrow\downarrow$ of the Fe(110), Fe(001) and Fe(111)
    monolayers with the nearest-neighbor distance of Cu fcc bulk
    $a=2.56$~\AA$\;$. The values are plotted as a function of the
    coordination.}
  \label{fig:strucCumuel}
  \caption{Magnetic moments and spin-orbit coupling constants
    $\lambda_{uu}^{l=2}$ for the spin-combinations $\uparrow\uparrow$
    and $\downarrow\downarrow$ of the Fe(110), Fe(001) and Fe(111)
    monolayers with the nearest-neighbor distance
    $a=2.4$~\AA$\;$. The values are plotted as a function of the coordination.}
  \label{fig:strucmuel}
  \caption{\protect NOLIMOKE spectrum of the different stripe structures for
    distances between the stripes varying from $h$ to $h+1.78$~\AA~as
    described in Fig.~\ref{fig:figchains}.}
  \label{fig:stripesnoli}
  \caption{Magnetic moment and spin-orbit coupling constant
    $\lambda_{uu}^{l=2}$ for the spin-combinations $\uparrow\uparrow$
    and $\downarrow\downarrow$ for the different stripe structures as
    a function of their ``interstripe'' distance $d$. }
  \label{fig:stripesmuel}
\end{figure}
%
\newpage
\psfig{file=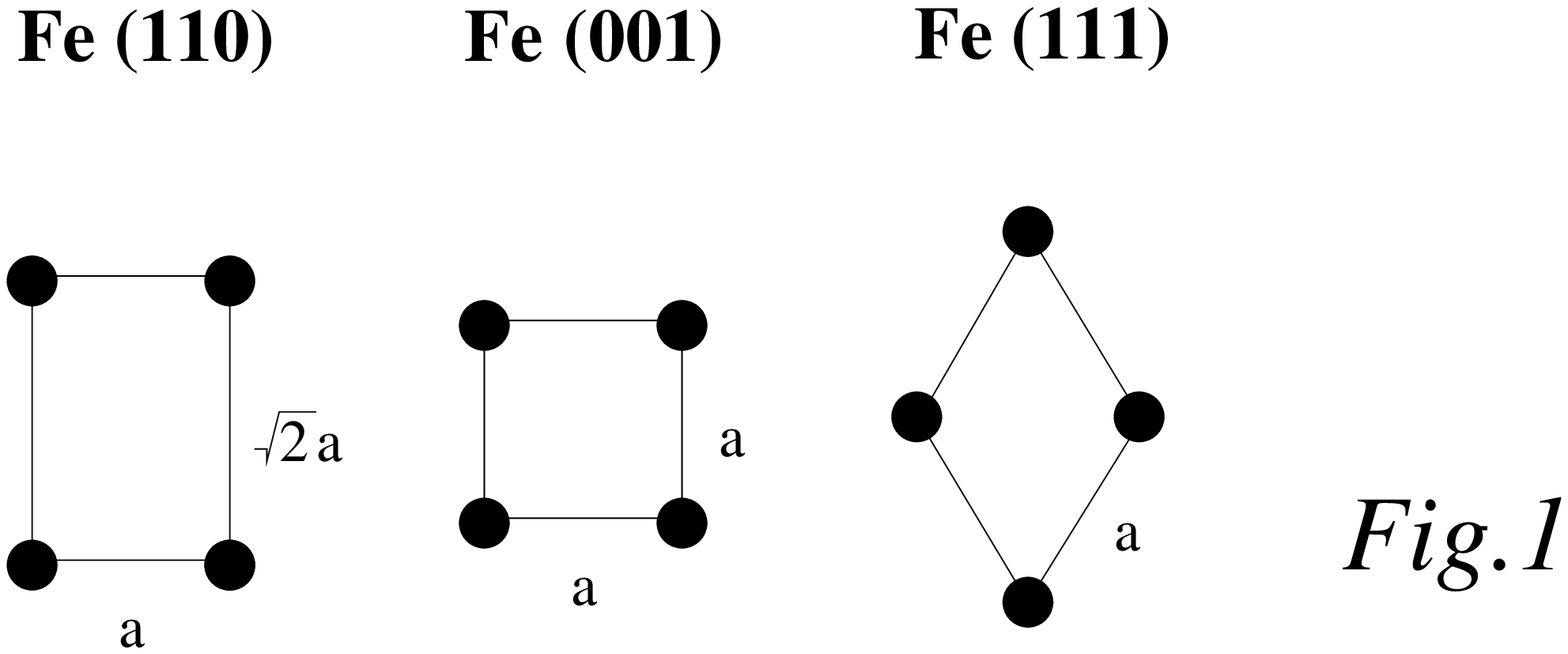,angle=-90,width=10cm}
\psfig{file=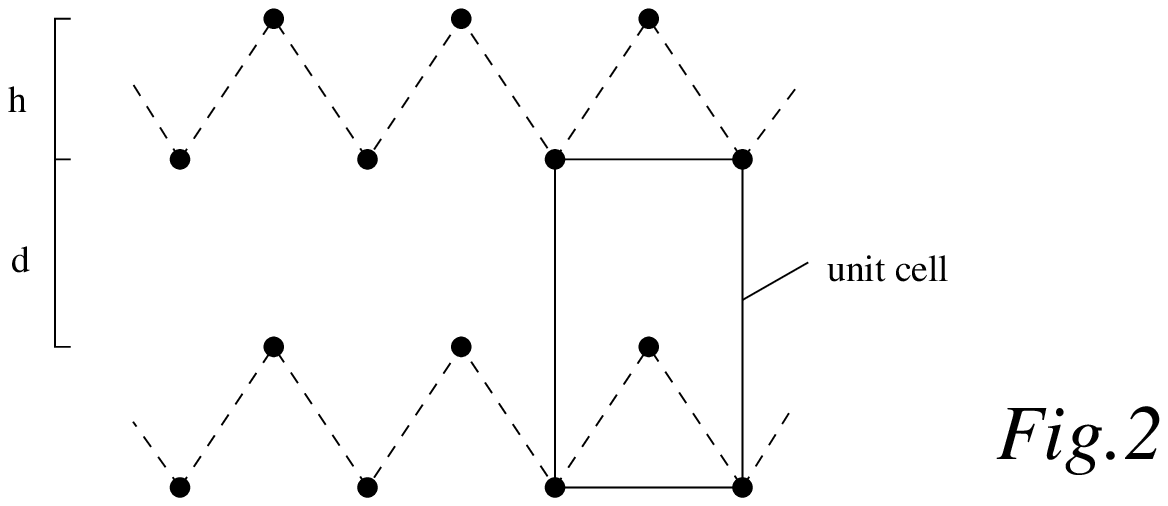,angle=-90,width=10cm}
\psfig{file=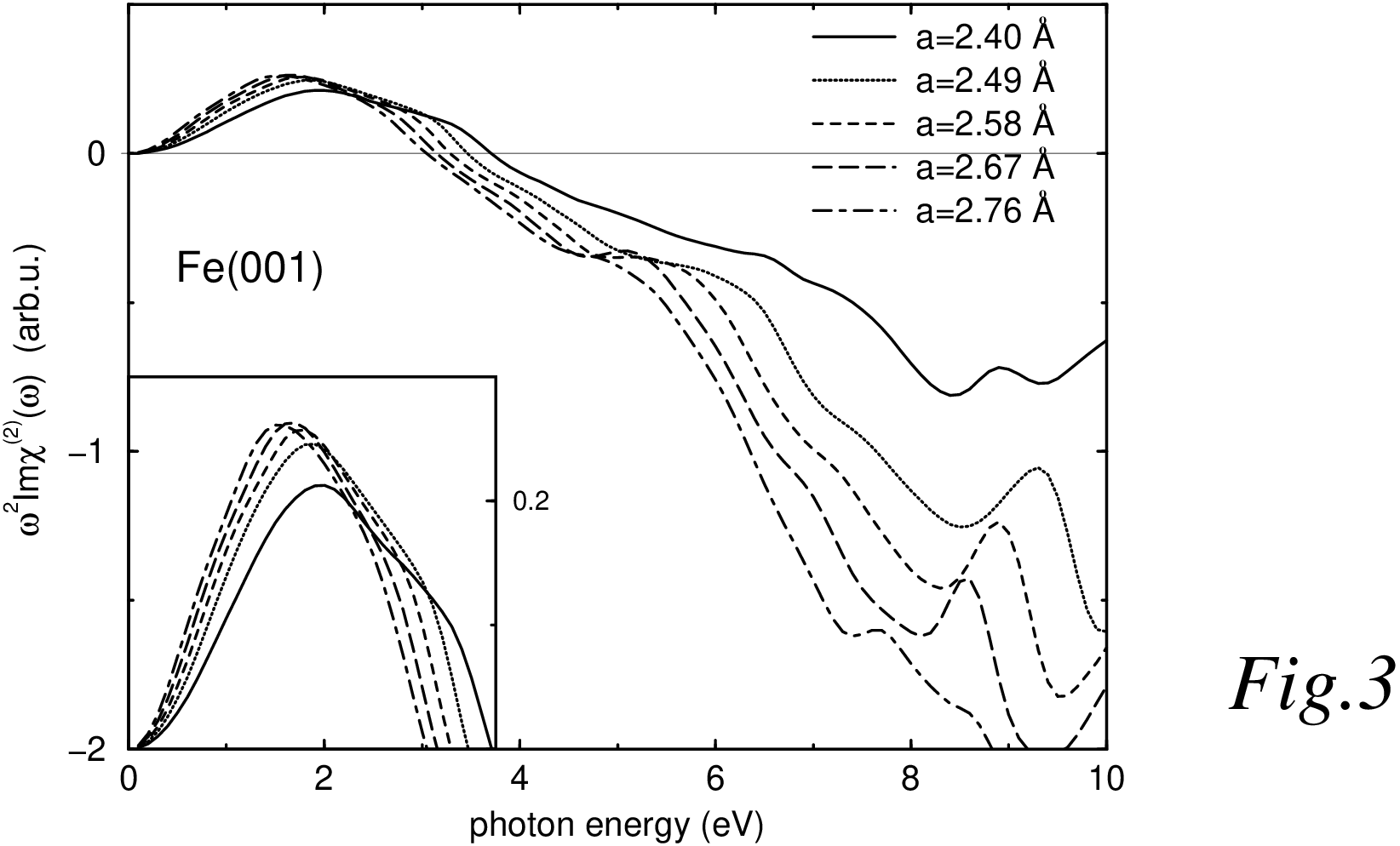,angle=-90,width=10cm}
\psfig{file=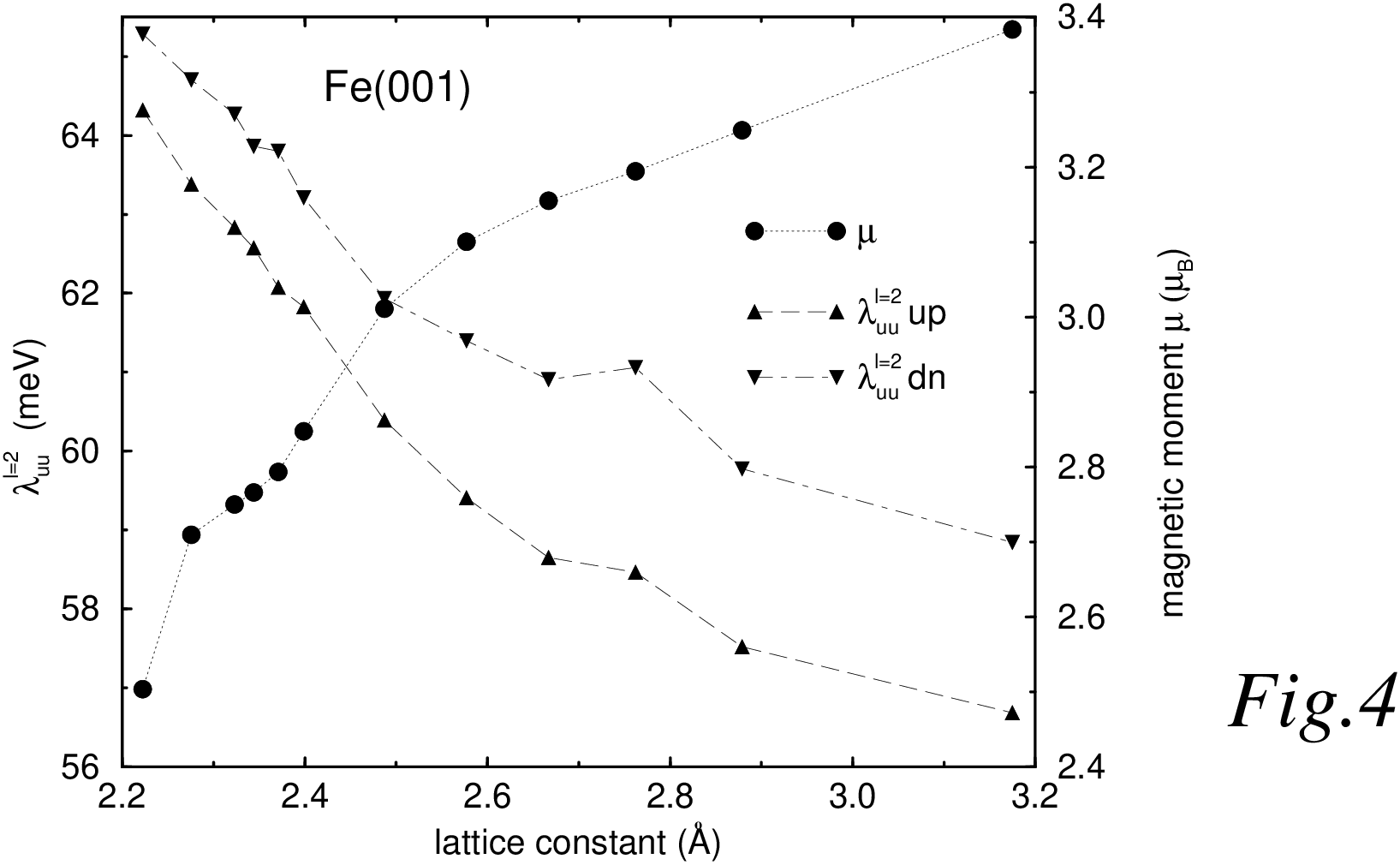,angle=-90,width=10cm}
\newpage
\psfig{file=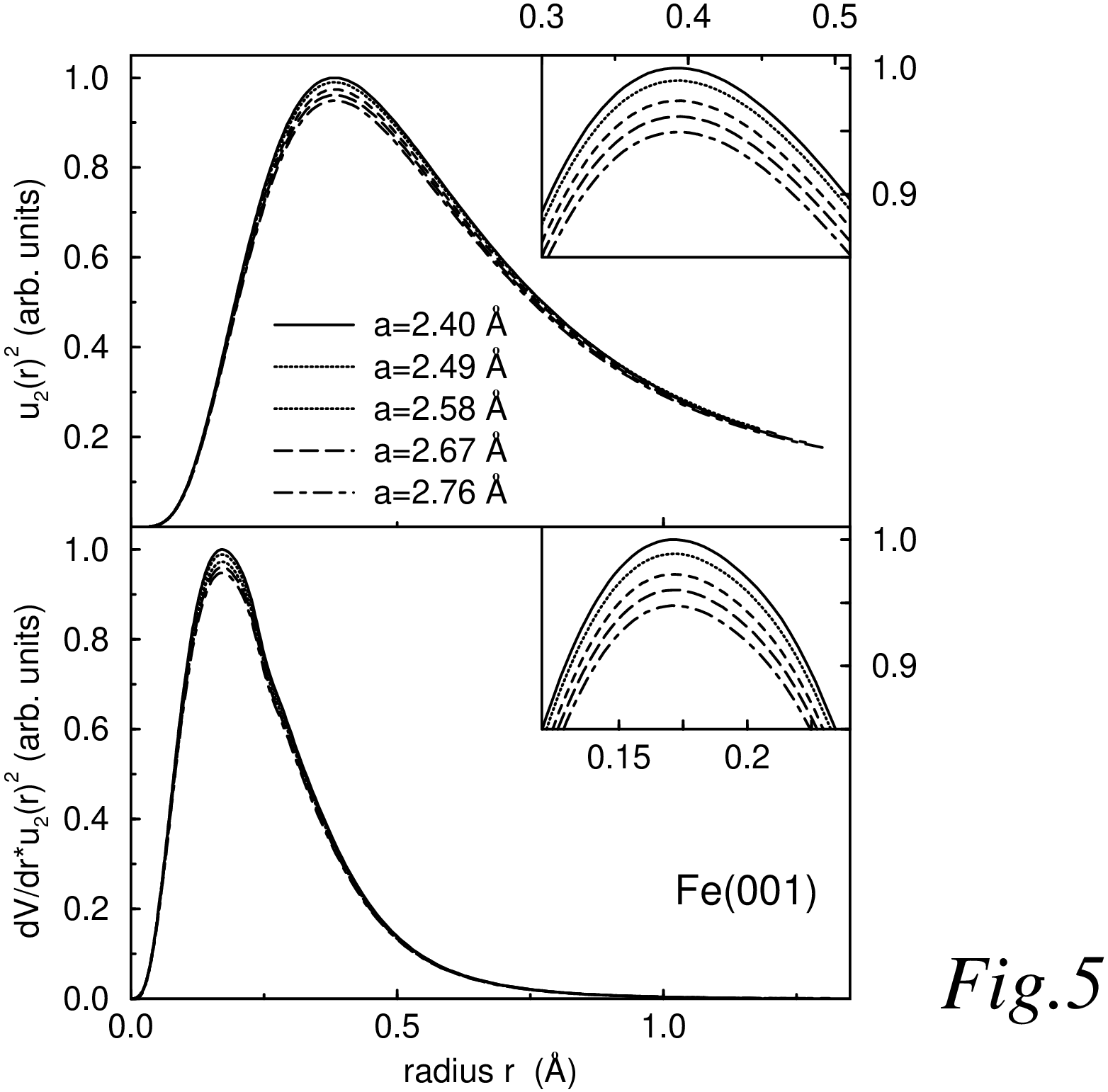,width=12cm}
\psfig{file=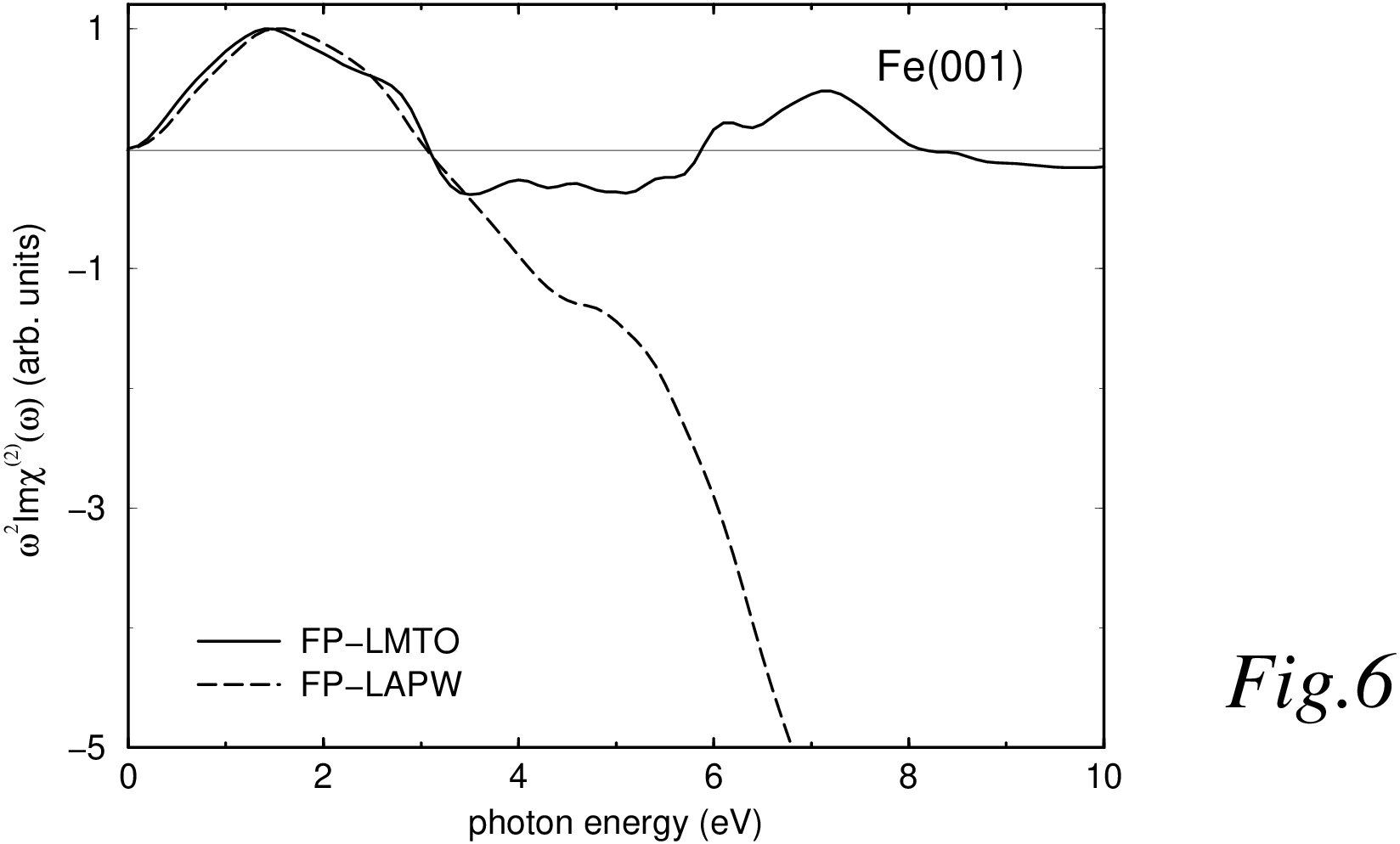,angle=-90,width=10cm}
\psfig{file=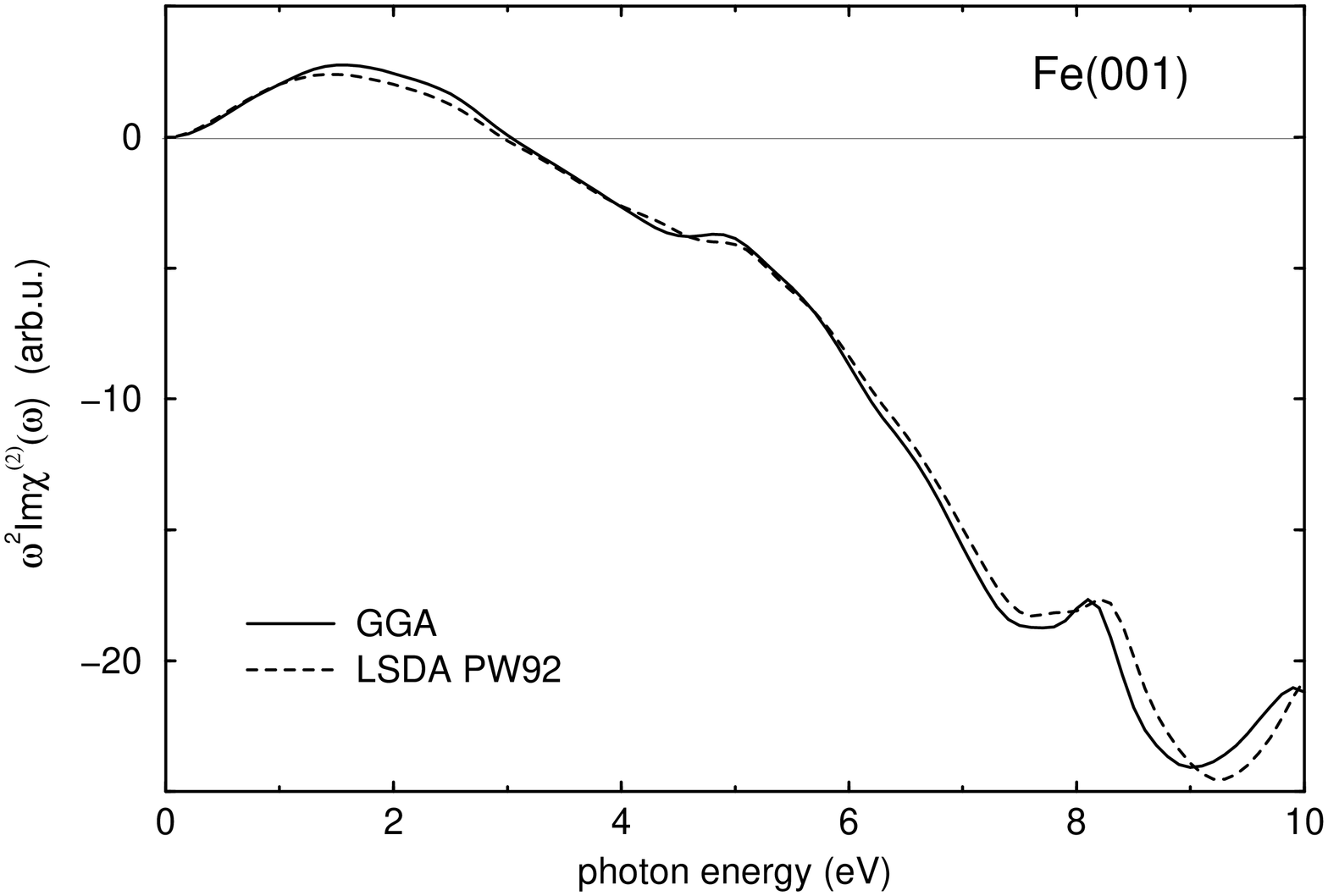,angle=-90,width=10cm}
\psfig{file=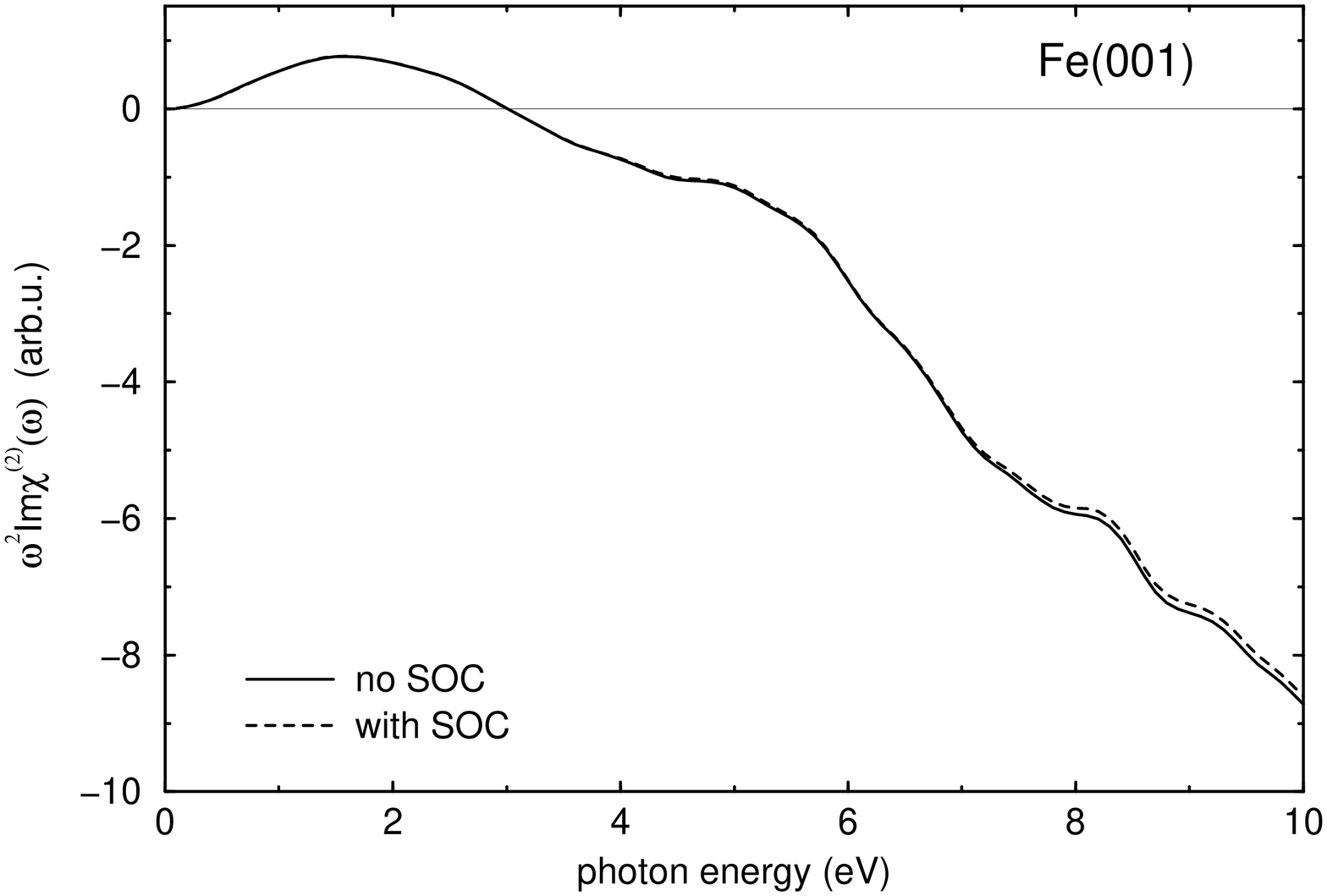,angle=-90,width=10cm}
\psfig{file=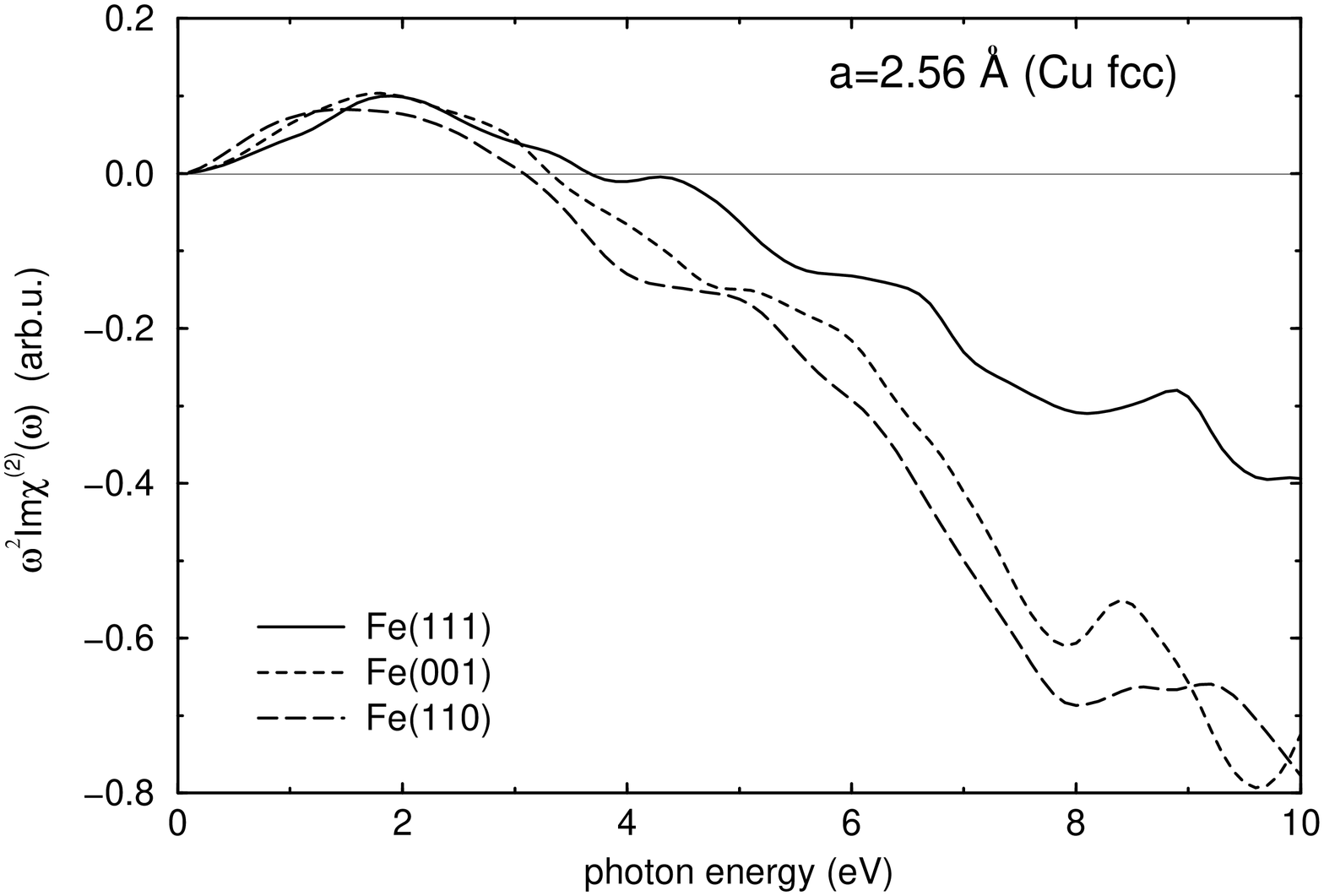,angle=-90,width=10cm}
\psfig{file=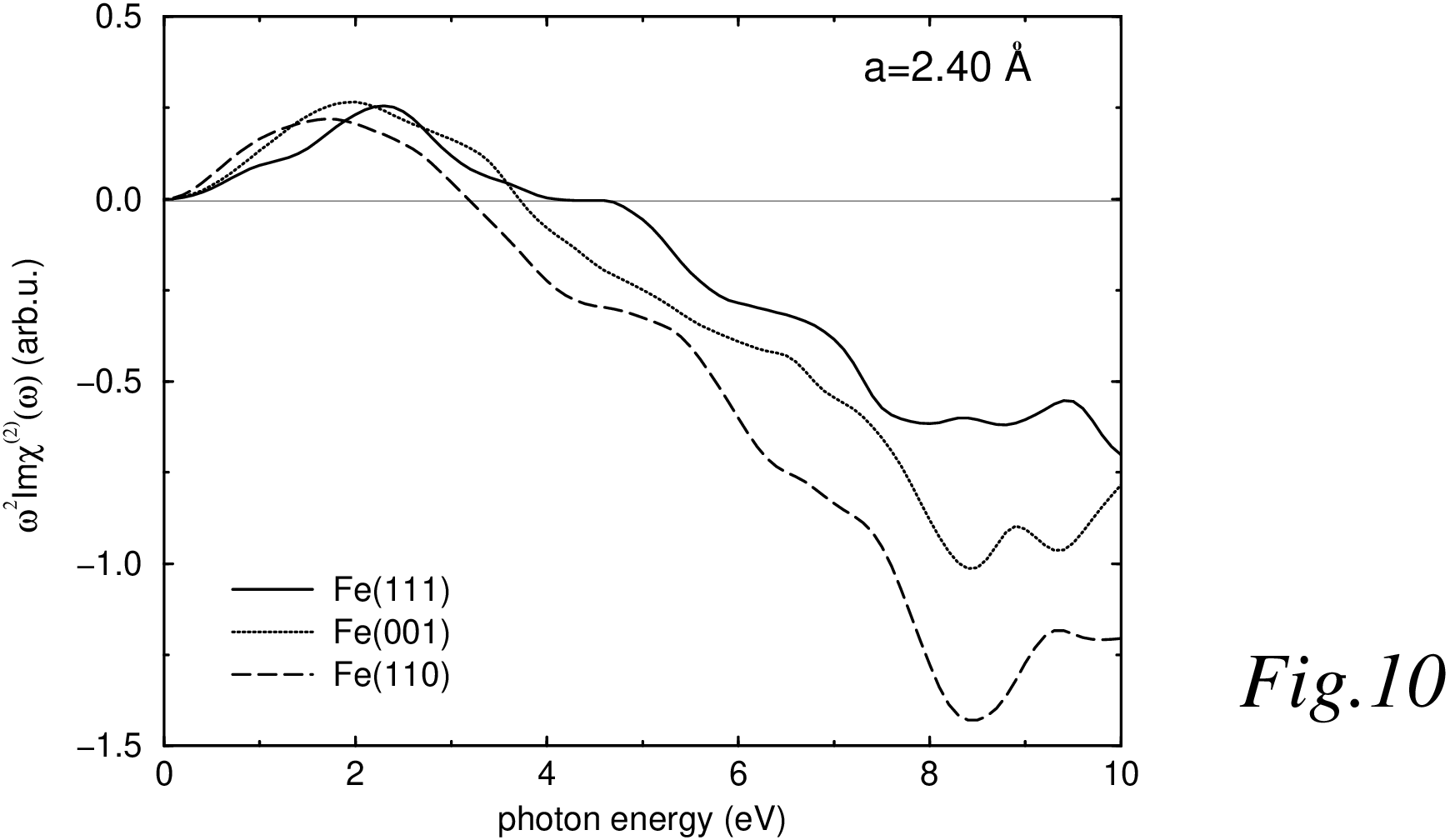,angle=-90,width=10cm}
\psfig{file=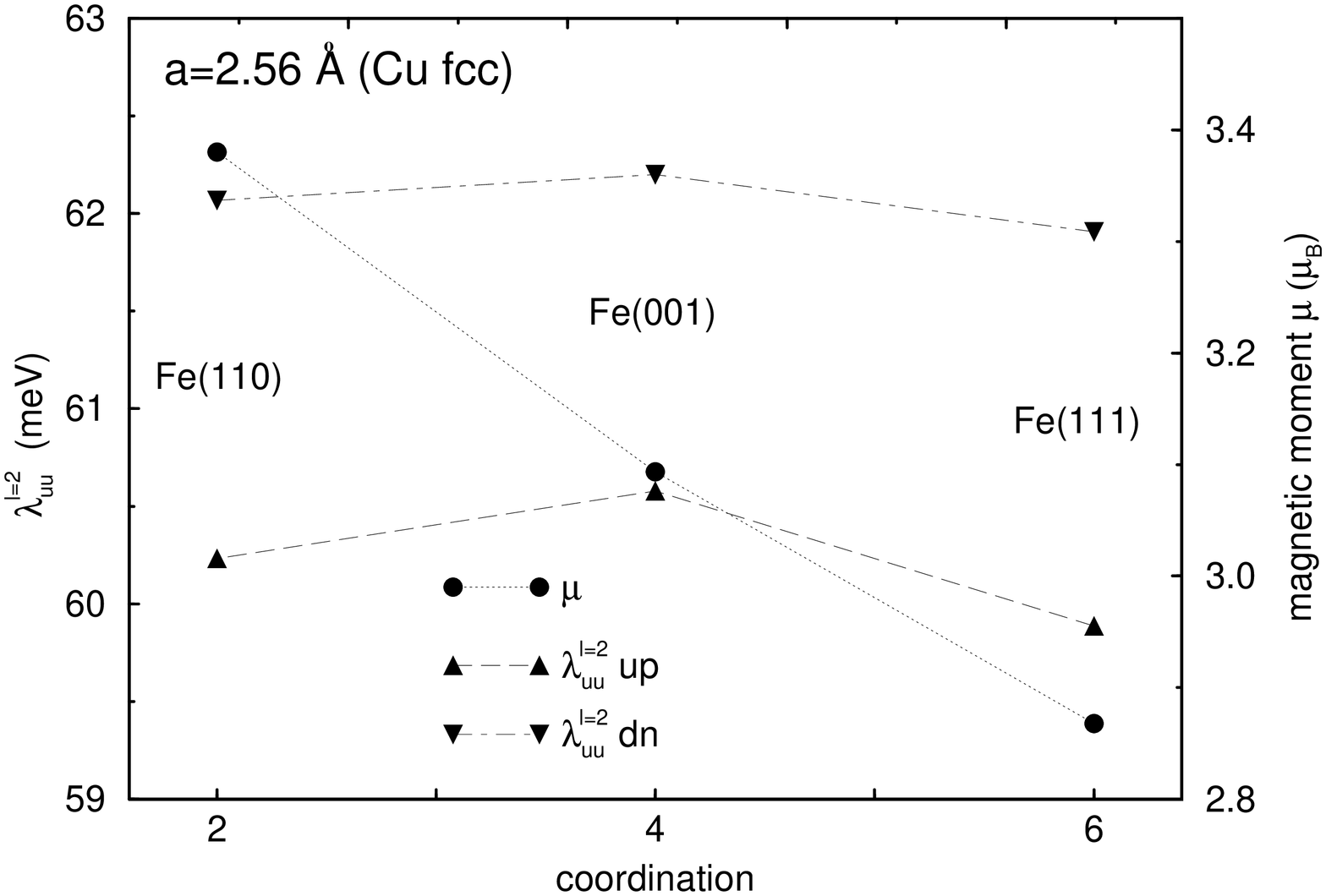,angle=-90,width=10cm}
\psfig{file=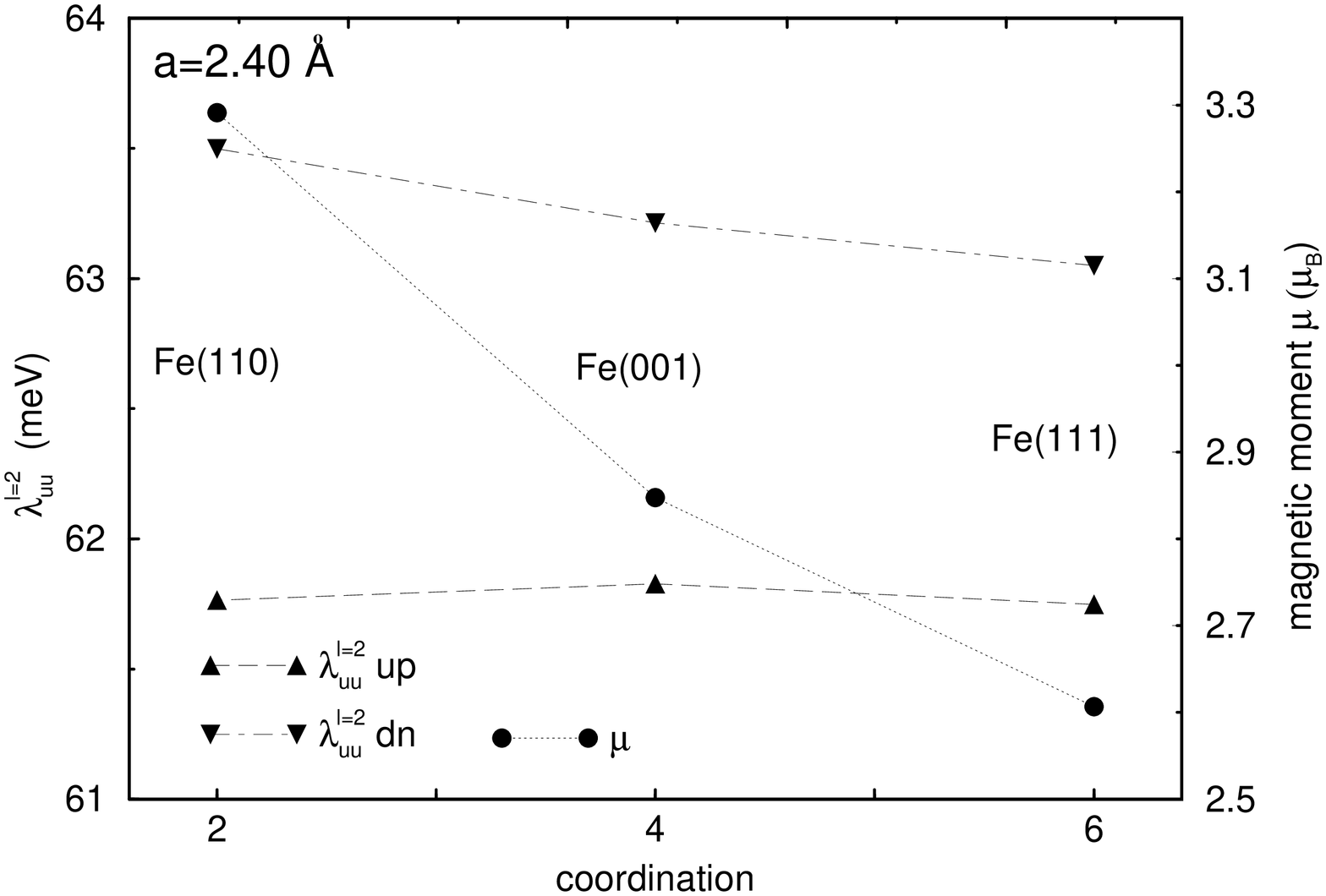,angle=-90,width=10cm}
\psfig{file=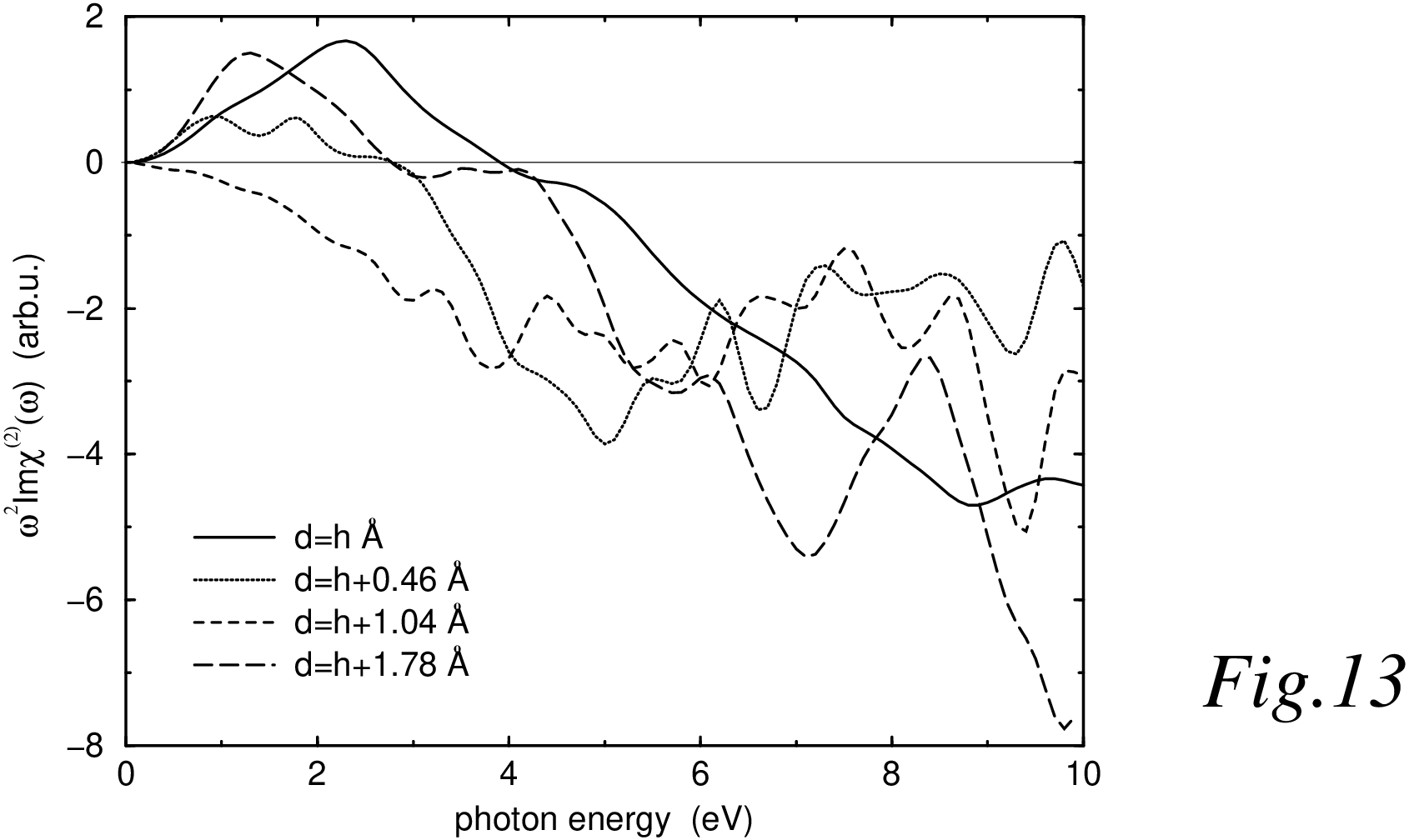,angle=-90,width=10cm}
\psfig{file=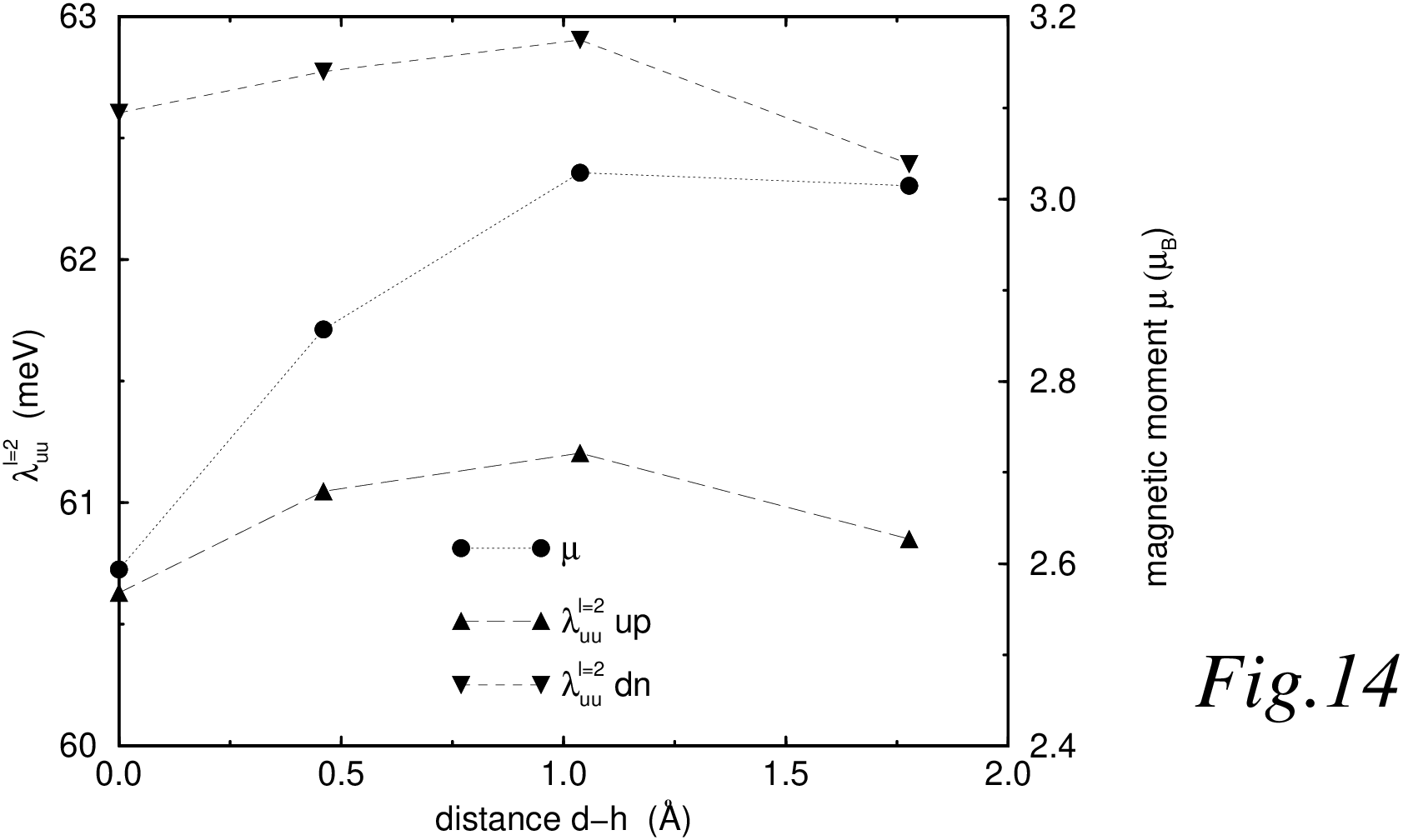,angle=-90,width=10cm}

%
%
\begin{table}[htbp]
  \begin{center}
    \leavevmode
    \caption{Values of the spin-orbit coupling constants
      $\lambda_{uu}^l$ and $\lambda_{\dot{u}\dot{u}}^l$ with
      $l=1,2$ for $\uparrow\uparrow$ and $\downarrow\downarrow$
      spin-combinations for the Fe(001) monolayer as a function of
      the lattice constants $a$ between 2.4 and 2.76~\AA$\;$.}  
    \label{tab:fe001}
    \caption{Values of the spin-orbit coupling constants
      $\lambda_{uu}^l$ and $\lambda_{\dot{u}\dot{u}}^l$ with
      $l=1,2$ for $\uparrow\uparrow$ and $\downarrow\downarrow$
      spin-combinations  for the Fe(111),
      Fe(001) and Fe(110) monolayer with the Cu fcc bulk nearest-neighbor
      distance $a=2.56$~\AA$\;$.}
    \label{tab:strucCu}
    \caption{Values of the spin-orbit coupling constants
      $\lambda_{uu}^l$ and $\lambda_{\dot{u}\dot{u}}^l$ with
      $l=1,2$ for $\uparrow\uparrow$ and $\downarrow\downarrow$
      spin-combinations for the Fe(111),
      Fe(001) and Fe(110) monolayer with the nearest-neighbor distance
      $a=2.4$~\AA$\;$.}
    \label{tab:struc}
    \caption{Values for the spin-orbit coupling constants
      $\lambda_{uu}^l$ and $\lambda_{\dot{u}\dot{u}}^l$ with
      $l=1,2$ for $\uparrow\uparrow$ and $\downarrow\downarrow$
      spin-combinations of the stripe
      structures as a function of the ``interstripe'' distance $d$.}
    \label{tab:stripes}
  \end{center}
\end{table}

%
%
{\bf\LARGE Tab. 1}\\[1ex]
\begin{tabular*}{13cm}{c|c|c|c|c|c|}
 [eV] & $a$=2.40~\AA &  $a$=2.49~\AA  & $a$=2.58~\AA & $a$=2.67~\AA &
 $a$=2.76~\AA \\[1ex] \hline 
$\qquad\lambda_{uu}^{l=1}$ $\uparrow\uparrow\qquad$ 
& 0.40789 & 0.33736 & 0.29471 & 0.26022 & 0.29714 \\
$\qquad\lambda_{uu}^{l=1}$ $\downarrow\downarrow\qquad$ 
& 0.41127 & 0.33583 & 0.29551 & 0.26304 & 0.31008 \\
$\qquad\lambda_{\dot{u}\dot{u}}^{l=1}$ $\uparrow\uparrow\qquad$ 
& 0.00082 & 0.00018 & 0.00061 & 0.00128 & 0.00601 \\
$\qquad\lambda_{\dot{u}\dot{u}}^{l=1}$ $\downarrow\downarrow\qquad$ 
& 0.00115 & 0.00041 & 0.00102 & 0.00186 & 0.00697 \\ \hline
$\qquad\lambda_{uu}^{l=2}$ $\uparrow\uparrow\qquad$ 
& 0.06182 & 0.06039 & 0.05941 & 0.05865 & 0.05909 \\
$\qquad\lambda_{uu}^{l=2}$ $\downarrow\downarrow\qquad$ 
& 0.06321 & 0.06193 & 0.06140 & 0.06091 & 0.06095 \\
$\qquad\lambda_{\dot{u}\dot{u}}^{l=2}$ $\uparrow\uparrow\qquad$ 
& 0.01372 & 0.01604 & 0.01868 & 0.02164 & 0.02459 \\
$\qquad\lambda_{\dot{u}\dot{u}}^{l=2}$ $\downarrow\downarrow\qquad$ 
& 0.01385 & 0.01621 & 0.01883 & 0.02175 & 0.02489 \\ \hline
\end{tabular*}\\[2ex]

%
%
%
{\bf\LARGE Tab. 2}\\[1ex]
\begin{tabular*}{11cm}{c|c|c|c|}
[eV] &$\quad$ Fe(111)$\quad$ &$\quad$ Fe(001)$\quad$ &$\quad$
Fe(110)$\quad$ \\[1ex] \hline
$\qquad\lambda_{uu}^{l=1}$ $\uparrow\uparrow\qquad$ 
& 0.34339 & 0.34822 & 0.34601 \\
$\qquad\lambda_{uu}^{l=1}$ $\downarrow\downarrow\qquad$ 
& 0.35045 & 0.35664 & 0.35519 \\
$\qquad\lambda_{u\dot{u}}^{l=1}$ $\uparrow\uparrow\qquad$ 
& 0.00244 & 0.00262 & 0.00252 \\
$\qquad\lambda_{u\dot{u}}^{l=1}$ $\downarrow\downarrow\qquad$ 
& 0.00299 & 0.00322 & 0.00316 \\ \hline
$\qquad\lambda_{uu}^{l=2}$ $\uparrow\uparrow\qquad$ 
& 0.05989 & 0.06058 & 0.06023 \\
$\qquad\lambda_{uu}^{l=2}$ $\downarrow\downarrow\qquad$ 
& 0.06191 & 0.06220 & 0.06207 \\
$\qquad\lambda_{u\dot{u}}^{l=2}$ $\uparrow\uparrow\qquad$ 
& 0.01807 & 0.01784 & 0.01790 \\
$\qquad\lambda_{u\dot{u}}^{l=2}$ $\downarrow\downarrow\qquad$ 
& 0.01818 & 0.01802 & 0.01802 \\ \hline
\end{tabular*}\\[2ex]

%
%
{\bf\LARGE Tab. 3}\\[1ex]
\begin{tabular*}{11cm}{c|c|c|c|}
[eV] &$\quad$ Fe(111)$\quad$&$\quad$ Fe(001)$\quad$ &$\quad$
 Fe(110)$\quad$ \\[1ex] \hline
$\qquad\lambda_{uu}^{l=1}$ $\uparrow\uparrow\qquad$ 
& 0.40746 & 0.40789 & 0.40900 \\
$\qquad\lambda_{uu}^{l=1}$ $\downarrow\downarrow\qquad$ 
& 0.41047 & 0.41127 & 0.41360 \\
$\qquad\lambda_{u\dot{u}}^{l=1}$ $\uparrow\uparrow\qquad$ 
& 0.00084 & 0.00082 & 0.00086 \\
$\qquad\lambda_{u\dot{u}}^{l=1}$ $\downarrow\downarrow\qquad$ 
& 0.00114 & 0.00115 & 0.00123 \\ \hline
$\qquad\lambda_{uu}^{l=2}$ $\uparrow\uparrow\qquad$ 
& 0.06175 & 0.06183 & 0.06176 \\
$\qquad\lambda_{uu}^{l=2}$ $\downarrow\downarrow\qquad$ 
& 0.06305 & 0.06321 & 0.06350 \\
$\qquad\lambda_{u\dot{u}}^{l=2}$ $\uparrow\uparrow\qquad$ 
& 0.01380 & 0.01372 & 0.01370 \\
$\qquad\lambda_{u\dot{u}}^{l=2}$ $\downarrow\downarrow\qquad$ 
& 0.01392 & 0.01385 & 0.01378 \\ \hline
\end{tabular*}\\[2ex]

%
%
\newpage
{\bf\LARGE Tab. 4}\\[1ex]
\begin{tabular*}{13cm}{c|c|c|c|c|}
[eV] & $d$=$h$ & $d$=$h+0.46$~\AA & $d$=$h+1.04$~\AA & $d$=$h+1.78$~\AA \\[1ex] \hline
$\qquad\lambda_{uu}^{l=1}$ $\uparrow\uparrow\qquad$ 
& 0.35653 & 0.38331 & 0.38857 & 0.38502 \\
$\qquad\lambda_{uu}^{l=1}$ $\downarrow\downarrow\qquad$ 
& 0.35388 & 0.38784 & 0.39421 & 0.39004 \\
$\qquad\lambda_{u\dot{u}}^{l=1}$ $\uparrow\uparrow\qquad$ 
& 0.00008 & 0.00127 & 0.00146 & 0.00130 \\
$\qquad\lambda_{u\dot{u}}^{l=1}$ $\downarrow\downarrow\qquad$ 
& 0.00022 & 0.00166 & 0.00189 & 0.00171 \\ \hline
$\qquad\lambda_{uu}^{l=2}$ $\uparrow\uparrow\qquad$ 
& 0.06063 & 0.06105 & 0.06120 & 0.06085 \\
$\qquad\lambda_{uu}^{l=2}$ $\downarrow\downarrow\qquad$ 
& 0.06261 & 0.06277 & 0.06290 & 0.06239 \\
$\qquad\lambda_{u\dot{u}}^{l=2}$ $\uparrow\uparrow\qquad$ 
& 0.01518 & 0.01507 & 0.01502 & 0.01504 \\
$\qquad\lambda_{u\dot{u}}^{l=2}$ $\downarrow\downarrow\qquad$ 
& 0.01523 & 0.01519 & 0.01514 & 0.01519q \\ \hline
\end{tabular*}\\[2ex]


\begin{thebibliography}{99}
\bibitem{coord}In general the fcc density of states is much more
  structured and asymmetric than the bcc density of states.
\end{thebibliography}
\end{document}